\documentclass[12pt]{article}

\usepackage[utf8]{inputenc}
\usepackage[T1]{fontenc}
\usepackage[ngerman, english]{babel}

\usepackage[]{geometry}
\usepackage{booktabs}

\usepackage[unicode]{hyperref}

\usepackage{subfig}
\usepackage{graphicx}

\usepackage{amsmath, amsfonts, amssymb, amsthm}
\usepackage{braket, stmaryrd}
\usepackage{tensor}

\let\bs\boldsymbol
\let\t\tensor
\let\p\partial
\def\frame{\mathfrak e}

\def\Fb{\bar F}				
\def\go{\gamma}				
\def\Ds{\nabla}				
\def\dd{\mathrm d}			
\def\order{O}

\newcommand{\braketP}[2]{(#1,#2)}
\newcommand{\braketM}[2]{\{#1,#2\}}

\newcommand{\bessel}[2]{L^{#1}_{#2}}

\DeclareMathOperator{\sgn}{sgn}		
\DeclareMathOperator{\diag}{diag}	
\DeclareMathOperator{\tr}{tr}		
\DeclareMathOperator{\adj}{adj} 	

\newcommand{\disc}[1]{\left\llbracket{#1}\right\rrbracket}
\newcommand{\discM}{\mathsf D}

\newcommand{\omlab}{\Omega}
\newcommand{\vecomlab}{\vec\Omega}

\usepackage{xcolor}

\usepackage{mdframed}
\newmdenv[
	hidealllines=true,
    backgroundcolor=black!05,
    innerleftmargin=0,
    innerrightmargin=0,
    innertopmargin=0.5\baselineskip,
    innerbottommargin=0.5\baselineskip
    skipbelow=0.5\baselineskip,
    skipabove=0.5\baselineskip
    ]{highlight}

\newcommand{\mnotex}[1]{}

\newcommand{\ptc}[1]{\mnotex{{\bf ptc:} {  #1}}}
\newcommand{\ptcr}[1]{\mnotex{{\bf ptc:} {\color{red} #1}}}
\newcommand{\tbm}[1]{\mnotex{{\bf tbm:} {\color{black} #1}}}
\newcommand{\tbmr}[1]{\mnotex{{\bf tbm:} {\color{red} #1}}}

\newcommand{\ptcheck}[1]{\ptc{checked on #1}}

\usepackage{microtype}

\usepackage[locale = US]{siunitx}
\DeclareSIUnit\year{yr}

\title{On the Influence of Earth’s Rotation on Light Propagation in Waveguides}
\date{\today}
\author{Thomas B. Mieling}

\begin{document}

\maketitle

\begin{abstract}

We analyse the influence of Earth’s rotation (both around its own axis and around the Sun) on the propagation of light in optical media.
This is done using both geometrical optics and a perturbative calculation based on Maxwell’s equations in rotating coordinates in flat spacetime.
Considering light propagation in cylindrical step-index waveguides in particular, the first order correction to electromagnetic modes is computed.
The calculation shows that Earth’s rotation causes a weak mode coupling, giving rise to sidebands, whose amplitudes are computed as well.
 The correction to the dispersion relation derived here allows to assess the anisotropy of light propagation due to Earth’s rotation.
 The linearisation of this result is found to agree numerically with a simple formula derived from geometrical optics.

\end{abstract}

\tableofcontents
\clearpage

\section{Introduction}
The problem of determining the influence of Earth’s rotation on optical phenomena was raised in the second half of the 19\textsuperscript{th} century and was first discussed using the theory of a luminiferous aether, see e.g.~\cite{Lorentz1887, Lorentz1923}.
Using an argument based on Huygen’s principle, Lorentz concluded that the effective refractive index $n_\mathrm{eff}$ of any medium is related to the ordinary refractive index $n$ by
\begin{equation}
	n_\mathrm{eff} \approx n - \vec m \cdot \vec v
	\,,
\end{equation}
cf.~\cite[p.~126ff]{Lorentz1887}, where $\vec v$ is the velocity of the supposed aether and $\vec m$ is the unit vector in the direction of light propagation. 

One of the main results of this paper is the derivation of an identical formula, by analysing the Maxwell’s equations in a rotating system in Minkowski spacetime. Here $\vec v$ is instead the negative linear velocity due to rotation of Earth, as measured from an inertial system at its center.
Moreover, we compute the first order corrections to both electric and magnetic field explicitly. In particular, we derive corrections to the  amplitudes of the modes and their sidebands, which are excited due to rotation.

The influence of Earth’s rotation (about its own axis) on light was first demonstrated experimentally using a Sagnac interferometer \cite{Pascoli2017}, where a beam of light is split into two, which traverse a ring interferometer in opposite sense of rotation. If the apparatus rotates, the different light rays take unequal times to complete a full circle, leading to an interference pattern whose fringes are displaced from the positions they would have if the apparatus were an inertial system.
Although Sagnac interpreted this effect as a proof for the existence of aether \cite{Sagnac1913,Sagnac1914}, Laue gave an explanation in the framework of Special Relativity \cite{Laue1920} and Langevin was able to explain the phenomenon in the language of General Relativity \cite{Langevin1921,Langevin1937}.

In a 2017 paper \cite{Hilweg_2017}, Hilweg et al.\ proposed an experiment to measure the effect of Earth’s gravitational potential on single photons. This paper included an estimate for the influence of Earth’s rotation, where the photon was modelled as a classical particle traversing a spooled waveguide. In a more recent paper \cite{Beig_2018}, the shift of the wave vector due to Earth’s gravitational field was calculated more accurately using Maxwell's equations in a post-Newtonian metric.

In this work, the effect of Earth's rotation is determined from first principles, using Maxwell’s equations in a rotating frame in Minkowski spacetime, thereby supplementing the well-known geometrical optics argument by more accurate wave optics calculations.

\subsection{Outline of this Work}

Classical propagation of light can be described either by geometrical optics or wave optics.
The latter method describes light by wave solutions of Maxwell's equations whereas geometrical optics is based on the eikonal equation, which can be understood as a limiting case of wave optics where the wavelength becomes infinitely short.

In Section~\ref{chapter:geometrical optics}, the influence of Earth’s rotation on the propagation of light is estimated using geometrical optics. Section~\ref{chapter:maxwell equations} is concerned with the derivation of a wave equation from Maxwell's equations in appropriate coordinates.
An approximate form of the equation is solved in Section~\ref{chapter:first approximation}. In Section \ref{s:further corrections}, it is shown that more accurate equations essentially lead to the same dispersion relation.
Finally, in Section~\ref{chapter:application} numerical examples of a concrete setup are provided and the effects of Earth's rotation around its own axis and the rotation about the Sun are compared.

\subsection{Model Assumptions}

In order for wave optics to be applicable, we must specify the full geometry of the medium of interest.
We consider cylindrical step index waveguides which consist of a cylindrical core of radius $a$ and refractive index $n_1$ surrounded by a cladding of diameter $a' \gg a$ and refractive index $n_2 < n_1$.
The dielectric will be assumed to be nonmagnetic and its field response will be assumed to be linear.
The dependence of the fields $\vec D$ and $\vec H$ on $\vec E$ and $\vec B$ is described covariantly by means of the \emph{optical metric}, introduced in \cite{Gordon1923} (not to be confused with the optical metric defined when studying null geodesics in stationary spacetimes).

We will neglect boundary effects which arise at the two ends of the waveguide and at the outer boundary of the cladding.
In this sense, the waveguide will be treated as infinitely long and the cladding as infinitely thick.

To describe Earth's rotation we use rigidly rotating coordinates in flat spacetime. The metric tensor will be approximated under the assumption that the rotational velocity is sufficiently slow compared to the speed of light.
Such a description applies both to Earth's rotation about its own axis and its orbit around the Sun (neglecting its orbital eccentricity).

\subsection{Conventions}

We use Heaviside-Lorentz units which, in our problem, coincide with Gaussian units, due to the absence of external charges and currents.
Furthermore, unless explicitly specified otherwise, we use units where the speed of light in vacuum is equal to one.
Compared to SI units, this means that
\begin{equation}
	c = \mu_0 = \varepsilon_0 = 1 \,.
\end{equation}
The signature of the metric tensor $g$ is
\begin{equation}
	\sgn(g) = (-, +, +, +)\,.
\end{equation}
The Einstein summation convention is used. Greek indices will range from $0$ to $4$ and Latin indices range from $1$ to $3$.

\section{Geometrical Optics}
\label{chapter:geometrical optics}
In this section we discuss the metric tensor used to describe rotating coordinate systems and introduce the optical metric to describe light propagation in rotating media.
A first estimate on the correction to the dispersion relation is derived in the framework of geometrical optics.

\subsection{The Metric Tensor in Born Coordinates}

It is convenient to use coordinates in which the waveguide is at rest.
In cylindrical Born coordinates $(t, \rho, \phi, z)$ the flat spacetime metric takes the form
\begin{equation}
	g =
	- (1- \Omega^2 \rho^2) dt^2 
	+ 2 \Omega \rho^2 dt d \phi
	+ d \rho^2
	+ \rho^2 d\phi^2
	+ dz ^2 \,.
\end{equation}
For slowly rotating systems, we may neglect terms quadratic in $v = \Omega \rho$. We thus arrive at the following metric in “rotating Cartesian coordinates” $(x^\mu) = (t,x,y,z)$:
\begin{equation}
	\label{eq:metric tensor}
	g = \t\eta{_\mu_\nu} dx^\mu dx^\nu + 2 v_i dt dx^i + \order(v^2)\,,
\end{equation}
where $v_i = \t\epsilon{_i_j_k} \Omega^j x^k$ is the local velocity field and $\vec \Omega$ is the angular velocity of the rotating system.
In the chosen coordinates, the inverse metric tensor is given by
\begin{equation}
	\t g{^\mu^\nu} = \t g{_\mu_\nu} + \order(\rho^2 \Omega^2) \,.
\end{equation}

In what follows, the components of any tensor will refer to the corotating coordinate system $x^\mu$.
Furthermore, $\omlab |\vec x|/c$ will be assumed to be sufficiently small, so that all terms quadratic in $\rho \omlab$ can be neglected.

Note that wave equations (in particular those for the electromagnetic field) also depend on derivatives of the metric. By expanding the metric tensor to order $v = \Omega \rho$, the wave equation is not expanded in powers of $v$ alone, but also in powers of $L \Omega$, where $L$ is any characteristic length scale of the problem: in our case these are the length $\ell$ and the radius $a$ of the waveguide, as well as the wavelength $\lambda$.
Thus, by neglecting terms of order $\order(v^2)$ in the components of the metric tensor, we also neglect terms of order $\order(a^2 \Omega^2)$, $\order(\ell^2 \Omega^2)$ and $\order(\lambda^2 \Omega^2)$ in the field equations.
Since the typical length scales of the dielectric are much smaller than $\rho$ (in our case, this is either Earth’s radius or its distance to the Sun), these error terms are expected to be negligible in all practical applications.
By slight abuse of notation, we will refer to all of these correction terms by $\order(\Omega^2)$ and thus formally expand all equations in powers of the (dimensionful) angular velocity $\Omega$ .

\subsection{The Optical Metric}

The (contravariant) optical metric tensor $\go$ of a linear dielectric with four-velocity vector field $u$ is defined as
\begin{equation}
	\label{eq:optical metric}
	\begin{aligned}
		\t\go{^\mu^\nu}
		&= \t g{^\mu^\nu} + (1- n^2) u^\mu u^\nu \,,
	\end{aligned}
\end{equation}
where $g$ is the spacetime metric and $n$ is the refractive index.
In the comoving coordinates the four-velocity of the co-rotating dielectric is given by
\begin{equation}
	u = \frac{\p}{\p x^0} \,,
	\qquad
	\Leftrightarrow
	\qquad
	(u^\mu) = (1, 0, 0, 0) \,.
\end{equation}
Using the velocity vector field $\vec v = \vec \Omega \times \vec x$, the components of $\go$ (in the comoving coordinates) take the form
\begin{equation}
	\label{eq:optical metric inverse components}
	\go^{0 0} = - n^2\,,
	\qquad
	\go^{0 i} = v^i\,,
	\qquad
	\go^{i j} = \delta^{ij} \,,
\end{equation}
where terms of order $\order(v^2)$ have been neglected.

\subsection{Geometrical Optics Approximation}

The eikonal equation
\begin{equation}
	\label{eq:eikonal equation}
	\go(\dd \psi, \dd \psi) = 0
	\,,
\end{equation}
may now be used to obtain a first estimate on the influence of Earth’s rotation on the propagation of light. One could try to solve  \eqref{eq:eikonal equation}  in a cylindrical waveguide, using appropriate conditions at the core-cladding interface, which we have not attempted to do. 
Instead, to simplify the problem, we look for plane wave solutions where the dispersion relation (which relates $\vec k^2$ and $\omega^2$) \emph{in the absence of rotation} reduces to $k = n \omega$, where $n$ is the effective refractive index obtained by solving the full Maxwell’s equations in the unperturbed case.
Writing
\begin{equation}
	\beta^2 = \vec k^2
	\,,
	\qquad
	\vec k = \beta \vec m
	\,,
\end{equation}
where $\vec m$ is the unit vector along the constant vector $\vec k$, and using the explicit form \eqref{eq:optical metric inverse components} of the optical metric, the eikonal equation \eqref{eq:eikonal equation} with $k = \dd \psi = \omega \dd t - k_i \dd x^i$ yields the quadratic equation
\begin{equation}
	-n^2 \omega^2 + \beta^2 - 2 \beta \omega \vec m \cdot \vec v = 0
	\,,
\end{equation}
where $v_\parallel = \vec m \cdot \vec v$.
Expanding the positive solution in powers of $v$ one obtains
\begin{equation}
	\beta = n \omega + \omega v_\parallel + \order(v^2)
	\,,
\end{equation}
from which one obtains the first order relative correction
\begin{highlight}
\begin{equation}
	\label{eq:geom:main result}
	\frac{\delta \beta}{\beta}
	\approx
	\frac{v_\parallel}{n}
	\,,
\end{equation}
\end{highlight}
which approximates unexpectedly well one of the main results of our work below.
As stated before, we require the dispersion relation to reproduce $\beta = n \omega$ in the unperturbed case, where $n$ is the \emph{effective} refractive index obtained from the exact solution of the unperturbed problem, which depends on the precise geometry of the waveguide.

Comparing the wavelengths in two parallel interferometer arms at different heights, we find that the rotationally induced phaseshift in a Mach-Zehnder interferometer is
\begin{equation}
	\Delta \Phi_\text{rot} = \omega \Omega A \cos \vartheta\,,
\end{equation}
where $A$ is the enclosed area and $\vartheta$ is the angle between the direction of motion and the direction of light propagation.
Comparing this with the gravitationally induced phase shift
\begin{equation}
	\Delta \Phi_\text{grav} = n \omega \mathrm g A\,,
\end{equation}
where $\mathrm g$ is the local gravitational acceleration (see Appendix \ref{appendix:gravitational field} for a derivation of this formula), we find that the rotational effect exceeds the gravitational one by a factor of up to $\Omega/(n \mathrm g) \approx 1.6 \times 10^3$ (depending on the angle $\vartheta$).

In the next sections, we will compute $\delta \beta$ using Maxwell's equations in the metric \eqref{eq:metric tensor}, which requires  specialisation to a fixed geometry.
We will consider an infinite cylindrical step index waveguide, for which an exact solution in the absence of rotation is known.
\ptcheck{5VII19 up to here}

\section{Maxwell's Equations in Rotating Coordinates}
\label{chapter:maxwell equations}
In this section, we formulate Maxwell's equations in the linearised Born metric \eqref{eq:metric tensor} and derive a wave equation for the electric and magnetic fields.

\subsection{Field Equations}

In manifestly covariant formulations of electrodynamics, the fields $\vec E$ and $\vec B$ are subsumed by a two-form $F$, and the fields $\vec D$ and $\vec H$ are combined to form the bivector $\Fb$.
In the absence of electromagnetic charges or currents, Maxwell's equations can be written in the form
\begin{equation}
	\mathrm d F = 0\,,
	\qquad
	  \delta \Fb = 0\,,
\end{equation}
where $\mathrm d$ denotes the exterior derivative and $\delta$ is the codifferential.
In local coordinates, these equations take the form
\begin{equation}
	\t\p{_[_\mu} \t F{_\nu_\rho_]} = 0\,,
	\qquad
	\t \nabla{_\nu} \t{\Fb}{^\mu^\nu} = 0\,,
\end{equation}
where the square brackets indicates antisymmetrisation and $\nabla$ denotes the Levi-Civita covariant derivative associated with the spacetime metric $g$.
For linear media of permittivity $\varepsilon$ and permeability $\mu$, the optical metric $\gamma$ allows to write the relation between $F$ and $\Fb$ in the concise form
\begin{equation}
	\label{eq:optical metric F F bar}
	\mu \t {\Fb}{^\alpha^\beta}
	= \t \go{^\alpha^\rho} \t\go{^\beta^\sigma} \t F{_\rho_\sigma}
	\,,
\end{equation}
which correctly reduces to $\t\Fb{^\alpha^\beta} = \t {F}{^\alpha^\beta}$ in vacuum.

We now discuss a “3+1 split” of these equations, which allows us to derive a wave equation for the electric and magnetic fields.

\subsection{Decomposition of the Field Strength Tensors}

Following \cite[eq.~(3.5)]{Beig_2018} we decompose the two-form $F$ and the bivector $\Fb$ as
\begin{equation}
	\begin{aligned}
		F
		&= + e \wedge \dd x^0
		+ \epsilon_{ijk} b^i \dd x^j \wedge \dd x^k \,,
		\\
		\Fb
		&= - d \wedge \frac{\p}{\p x^0}
		+ \epsilon^{ijk} h_i \, \frac{\p}{\p x^j} \wedge \frac{\p}{\p x^k} \,,
	\end{aligned}
\end{equation}
where $e$ and $h$ are spatial one-forms and $b$ and $d$ (not to be confused with the differential operator $\mathrm d$) are spatial vectors.
More explicitly, this means
\begin{equation}
	\label{eq:field strength decomposition}
	\begin{aligned}
		\t e{_i}
			&= \t F{_i_0} \,,
		\qquad &
		\t b{^i}
			&= \frac{1}{2} \t \epsilon{^i^j^k} \t F{_j_k} \,,
		\\
		\t d{^i}
			&= \t {\Fb}{^0^i} \,,
		\qquad &
		\t h{_i}
			&= \frac{1}{2} \t \epsilon{_i_j_k} \t {\Fb}{^j^k} \,,
	\end{aligned}
\end{equation}
where, as already pointed out, all field components $F_{\mu\nu}$ and $\Fb^{\mu\nu}$ refer to the corotating coordinate system.

Using \eqref{eq:optical metric F F bar}, we may express the $d$ and $h$ fields in terms of the $e$ and $b$ fields as
\begin{equation}
	\label{eq:transformation matter fields}
	\begin{aligned}
		\mu d^i &= n^2 e_i - \t \epsilon{^i^j^k} v^j b^k \,,
		\\
		\mu h_i &= b^i - \t \epsilon{_i_j_k} v^j e_k \,,
	\end{aligned}
\end{equation}
where terms of order $\order(v^2)$ have been neglected.

\subsection{Decomposition of the Field Equations}

From \cite[eqns.~(3.12) and (3.15)]{Beig_2018} we have the following decomposition of the field equations
\begin{equation}
	\label{eq:maxwell}
	\begin{aligned}
		\dot b^i + \t\epsilon{^i^j^k} \partial_j e_k &= 0 \,,\\
		\dot d^i - \t\epsilon{^i^j^k} \partial_j h_k &= 0 \,,\\
		\partial_i b^i &= 0 \,,\\
		\partial_i d^i &= 0 \,.\\
	\end{aligned}
\end{equation}
We need not make any distinction between upper and lower indices since they are manipulated with the spatial metric $\delta_{ij}$.
Since $\vec b$ and $\vec d$ have a vanishing spatial divergence, it is natural to work with these two fields and express $\vec e$ and $\vec h$ in terms of them.
From \eqref{eq:transformation matter fields} one finds to order $\order(v)$
\begin{highlight}
\begin{equation}
	\label{eq:rotation:matter fields}
	\begin{aligned}
		n^2 \vec e &= \mu \vec d + \vec v \times \vec b \,,\\
		n^2 \vec h &= \epsilon \vec b - \vec v \times \vec d \,,
	\end{aligned}
\end{equation}
\end{highlight}
where $n^2 = \epsilon \mu$ was used.

We note that for $n \neq 1$, we may set $\vec u := - \vec v/(n^2 -1)$ to obtain
\begin{equation}
	\begin{aligned}
		\vec d &= \epsilon \vec e \;+ (n^2 - 1) \vec u \times \vec h\,,\\
		\vec b &= \mu \vec h - (n^2 - 1) \vec u \times \vec e\,,
	\end{aligned}
\end{equation}
which coincides with the relations between $\vec e$, $\vec b$ and $\vec d$, $\vec h$ in a medium which moves with velocity $\vec u$ relative to an inertial system \cite[p.~329, eqns.\ (76,10) and (76,11)]{Landau_08}.\ptcr{this is perplexing...}
This is due to the special form of the metric perturbation, which allows us to write
\begin{equation}
	\t \go{^\mu^\nu} =
		\t \eta{^\mu^\nu}
		+ \t {\delta g}{^\mu^\nu}
		+ (1-n)^2 u^\mu u^\nu
		=
		\t \eta{^\mu^\nu}
		+ (1-n)^2 \tilde u^\mu \tilde u^\nu
		+ \order(v^2)
		\,,
\end{equation}
where $(u^\mu)$ describes a dielectric at rest and $(\tilde u^\mu)$ describes slow movement of the velocity given above. However, this equivalence does not extend to higher powers of $v$.

Substituting \eqref{eq:rotation:matter fields} into \eqref{eq:maxwell}, we obtain the following field equations for $\vec d$ and $\vec b$.
\begin{equation}	
	\label{eq:rotation:maxwell covariant}
	\begin{aligned}
		n^2 \dot{\vec b}
		+ \mu \vec \nabla \times \vec d
		- [\vec v, \vec b] &= 0\,,
		\\
		n^2 \dot{\vec d}
		- \epsilon \vec \nabla \times \vec b
		- [\vec v, \vec d] &= 0\,,
		\\
		\vec \nabla \cdot \vec b &= 0\,,
		\\
		\vec \nabla \cdot \vec d &= 0\,,
	\end{aligned}
\end{equation}
where $[\vec v, \vec w]$ denotes the commutator (Lie bracket) of the vector fields $\vec v$ and $\vec w$.
\ptcheck{15III19 together}

\subsection{Electric-Magnetic Symmetry}

Note that the field equations are invariant under the following transformation among $\vec b$ and $\vec d$:
\begin{equation}
	\label{eq:symmetry B,D}
	\vec b \mapsto + \frac{\mu}{n} \vec d \,,
	\qquad
	\vec d \mapsto - \frac{\varepsilon}{n} \vec  b\,,
\end{equation}
which induces the following transformation of the fields $\vec e$ and $\vec h$:
\begin{equation}
	\label{eq:symmetry H,E}
	\vec h \mapsto + \frac{\varepsilon}{n} \vec e \,,
	\qquad
	\vec e \mapsto - \frac{\mu}{n} \vec h \,.
\end{equation}

\subsection{The Wave Equation for the Electromagnetic Field}
\label{s:EM wave equation}

The field equations \eqref{eq:rotation:maxwell covariant} imply the following wave equations for the electromagnetic field:
\begin{equation}
	n^2 (
		n^2 \ddot{\vec b}
		- \Delta \vec b
	)
	= 2 \nabla_{\vec v}\, (n^2 \dot{\vec b})
	- 2 \mu [
		(\vec \Omega \cdot \vec \nabla) \vec d - \vec \nabla (\vec \Omega \cdot \vec d)
	]
	+ \order(\Omega^2)
	\,.
\end{equation}
A derivation of these equations is given in Appendix~\ref{a:EM wave equation}.

To get some insight into the relative magnitudes of the two source terms, let us estimate spatial derivatives of the fields by the inverse wavelength $1/\lambda$. Since $\vec v$ is of the order $\Omega R$, we estimate that the first order differential operator is suppressed relative to the second order operator by the factor $\lambda/R$, which is clearly negligible.
In \ref{s:correction:wavelength}, we show that this term does not modify the main result obtained without it.

Furthermore, let us note that the directional derivative along $\vec v$ depends on the position in the dielectric.
Decomposing the radius vector $\vec x$ as $\vec x = \vec R + \vec r$, where $\vec R$ points from centre of rotation to one end of the dielectric, and choosing $z$ to be the distance along the symmetry axis of the waveguide, the coordinate ranges inside the core are
\begin{equation}
	x^2 + y^2 \leq a^2
	\qquad\text{and}\qquad
	0 \leq z \leq \ell
	\,.
\end{equation}
Neglecting the dependence of $\vec v$ on $x$ and $y$ is expected to produce relative errors of order $\order(a/R)$, while those resulting from neglecting its $z$ dependence are expected to be
\begin{equation}
	\order(\ell/R) \gg \order(a/R)
	\,,
\end{equation}
since the length of a waveguide is typically many orders of magnitude larger than its diameter.
As a first approximation we will use $\vec v \approx \vec V$, where
\begin{equation}
	\vec V = \vec \Omega \times \vec R
	\,,
\end{equation}
and in section \ref{s:correction:length} we show that corrections of relative order $\ell/R$ have no influence on the dispersion relation.

Instead of considering the full equations
\begin{equation}
	\label{eq:wave:full}
	\begin{aligned}
		\epsilon (
			n^2 \ddot{\vec b}
			- \Delta \vec b
		)
		&= 2 \epsilon\; \nabla_{\vec v}\, \dot{\vec b}
		- 2 [
			(\vec \Omega \cdot \vec \nabla) \vec d - \vec \nabla (\vec \Omega \cdot \vec d)
		]
		\,,
		\\
		\mu (
			n^2 \ddot{\vec d}
			- \Delta \vec d
		)
		&= 2 \mu \nabla_{\vec v}\, \dot{\vec d}
		- 2 [
			(\vec \Omega \cdot \vec \nabla) \vec b - \vec \nabla (\vec \Omega \cdot \vec b)
		]
		\,.
	\end{aligned}
\end{equation}
we may instead consider the simplified wave equations
\begin{equation}
	\label{eq:wave:approximate}
	\begin{aligned}
		n^2 \ddot{\vec b} - \Delta \vec b
		&= 2  \nabla_{\vec V}\, \dot{\vec b}
		\,,
		\\
		n^2 \ddot{\vec d} - \Delta \vec d
		&= 2 \nabla_{\vec V}\, \dot{\vec d}
		\,,
	\end{aligned}
\end{equation}
which yield the same dispersion relation up to terms of order $a/R$.

\section{Maxwell’s Equations in Cylindrical Waveguides}
In this section, we give an outline of the calculations in the framework of wave optics and review the solution to the unperturbed problem.

\subsection{Outline of the Calculation}
We seek solutions to Maxwell's equations inside and outside an infinite cylinder of radius $a$, where all field components depend on time $t$ and $z$ (distance along the symmetry axis) as $e^{i(\omega t - \beta z)}$.
All calculations are performed in corotating cylindrical coordinates $(t, r, \theta, z)$.
We choose as “fundamental fields”
\begin{equation}
	f =
	\begin{pmatrix}
		D^z\\
		B^z
	\end{pmatrix}
	\,,
\end{equation}
which determine all other field components uniquely.
Symbolically, we write $F = F(f; \beta, \omega)$, where $F$ is an abbreviation for the collection of fields ($\vec E, \vec D, \vec B, \vec H)$.
Due to the linearity of the medium, $F$ depends linearly on $f$, which satisfies a wave equation of the form
\begin{equation}
	n^2 \ddot f - \Delta f = g[f],
\end{equation}
where $g$ is a linear differential operator, which vanishes in the unperturbed case.

At the core-cladding interface ($r = a$), the field components $D^r, E^\theta, E^z, B^r, H^\theta$ and $H^z$ are required to be continuous.
As only four of these conditions are linearly independent, it suffices to require the continuity of  $D^r, E^z, B^r$ and $H^z$, for which we symbolically write
\begin{equation}
	\disc{F(f; \beta, \omega)} = 0\,,
\end{equation}
where $\disc{\cdot}$ is a linear measure of the jump height.

The problem at hand thus reduces to a wave equation for $f$ on two disjoint domains (core and cladding), subject to the linear continuity constraint at the interface $r = a$.


\subsubsection{Unperturbed Case}

In the unperturbed case, where $g$ vanishes, we consider the excitation of a single mode, i.e.\
\begin{equation}
	f^{(0)}(t,r,\theta,z) = h(r; \beta, \omega) e^{i (\omega t - \beta z + m \theta)}.
\end{equation}
The two-component radial function $h(r; \beta, \omega)$ has four (complex) degrees of freedom, since it satisfies a homogeneous ordinary differential equation (ODE) of second order on two disjoint domains (core and cladding) and is required to be regular at the origin and to decay at large distances.
We choose linear coordinates $\bs \alpha$ on the solution space and write $h = h_{\bs \alpha}$, such that the four linearly independent continuity conditions are equivalent to a matrix equation of the form
\begin{equation}
	M^{(0)}(\beta, \omega) \bs \alpha = 0\,.
\end{equation}
To obtain a nontrivial solution ($\bs \alpha \neq 0$) one is led to the equation
\begin{equation}
	\det M^{(0)}(\beta, \omega) = 0\,,
\end{equation}
which provides the dispersion relation linking $\beta$ and $\omega$.
Finally, the coefficients $\bs \alpha$ are chosen to be in the kernel of $M^{(0)}$. An explicit calculation shows that this kernel is one-dimensional.

\subsubsection{First Order Corrections}

To describe linear corrections due to Earth’s rotation, we write
\begin{equation}
	f(t, r, \theta, z)
	= \left[
	h_{\bs \alpha}(r; \beta, \omega) + f^{(1)}(r, \theta; \beta, \omega)
	\right]e^{i(\omega t - \beta z + m \theta)}
	\,,
\end{equation}
where $f^{(1)}$ also depends on $\theta$ due to the angular dependence of $g$.
Decomposing $f^{(1)}$ into a Fourier series in the angular variable, every coefficient function $f^{(1)}_k$ satisfies an inhomogeneous ODE of second order, subject to the same constraints as above, and is thus an element of a four-dimensional (complex) solution space.
Due to the linear independence of the angular modes, continuity conditions hold for all modes separately.
Further analysis requires a distinction between the main mode (no further dependence on $\theta$, so $k = 0$) and the sidebands ($k \neq 0$).

For the main mode, $f^{(1)}_0$ can be chosen to be any particular solution to the ODE, since a homogeneous part can be removed by redefining $\bs \alpha$. Thus, the radial function of the main mode depends linearly on $\bs \alpha$, so the continuity condition of this mode is again equivalent to a homogeneous matrix equation
\begin{equation}
	\left( M^{(0)}(\beta, \omega) + M^{(1)}(\beta, \omega)\right) \bs \alpha = 0\,,
\end{equation}
which determines the first order correction to the dispersion relation.

For the sidebands, the radial functions $f^{(1)}_k$ satisfy inhomogeneous ODEs, whose inhomogeneities are linear in the coefficients $\bs \alpha$.
Parameterising homogeneous solutions linearly with parameters $\bs \chi_k$, $f^{(1)}_k$ becomes a function linear in $\bs \alpha$ and $\bs \chi_k$, so the continuity condition is equivalent to an inhomogeneous matrix equation of the form
\begin{equation}
	M_k \bs \chi_k = N_k \bs \alpha \,,
\end{equation}
whose solution   $\bs \chi_k$ turns out to be unique.

In the next section, we shall review the unperturbed solution in light of this structure, and in later sections we will carry out the perturbational calculation, following the outline discussed here.

\subsection{Unperturbed Solution}
In this section, we review the solutions of the problem when $\omlab=0$, using the notation of \cite{Beig_2018} with $\phi = 0$ (i.e. no gravitational potential) and $c = \mu_0 = \epsilon_0 = 1$.

Since the dielectric is assumed to be nonmagnetic, we have $\mu = 1$ and thus $n^2 = \epsilon$.
The relation between $\vec D, \vec H$ and $\vec E, \vec B$ are thus
\ptc{this equation is consistent with mine if $\epsilon_0=1=\mu_0$ which, according to TM, is the definition of Gaussian units}
\begin{equation}
	\label{eq:macroscopic fields unperturbed}
	\begin{aligned}
		\vec D
			&= n^2 \vec E \,,\qquad
		\vec B
			&= \vec H \,.
	\end{aligned}
\end{equation}
The refractive index is assumed to be piecewise constant, namely
\begin{equation}
	n(r) =
	\begin{cases}
		n_1\,,	&	r < a \,,\\
		n_2\,,	&	r > a \,,
	\end{cases}
\end{equation}
where $a$ denotes the radius of the dielectric.

We separate the dependence of the fields on $t, z$ and $\theta$ via
\ptc{watch out, some equations in the published paper have $-m\theta$, but they are inconsistent; the consistent setup is with the signs here...}
\begin{equation}
	\begin{aligned}
		\vec E(t, \vec r) = \vec E(r) e^{i(\omega t- \beta z + m \theta)},\qquad
		\vec B(t, \vec r) = \vec B(r) e^{i(\omega t- \beta z + m \theta)}.
	\end{aligned}
\end{equation}
In the absence of a gravitational potential ($\phi = 0$), the constants $U$ and $W$ defined in \cite[eqns.~(4.17, 4.18)]{Beig_2018} reduce to
\ptc{this assumes no gravitational potential, so $\psi\equiv 1$}
\begin{equation}
	\begin{aligned}
		U^2 &= a^2 (n_1^2 \omega^2 - \beta^2)\,,\\
		W^2 &= a^2 (\beta^2 -n_2^2 \omega^2)\,,
	\end{aligned}
\end{equation}
with $U$ and $W$ positive.
The function
\begin{equation}
	f(r) =
	\begin{cases}
		J_m(U r/a)/J_m(U)\,,	&	r<a \,,\\
		K_m(W r/a)/K_m(W)\,,	&	r>a \,,\\
	\end{cases}
\end{equation}
is the continuous solution of the equations
\begin{equation}
	\begin{aligned}
		f''(r) + \frac{1}{r} f'(r) + \left( \frac{U^2}{a^2} - \frac{m^2}{r^2} \right)f(r) &= 0\qquad \text{for }r<a \,,\\
		f''(r) + \frac{1}{r} f'(r) - \left( \frac{W^2}{a^2} + \frac{m^2}{r^2} \right)f(r) &= 0\qquad \text{for }r>a \,,
	\end{aligned}
\end{equation}
which is bounded when $r \to 0 $ and has finite total energy.
The $z$ components of the fields are then given by
\begin{equation}
	\begin{aligned}
		E^z(r) &= A f(r) \,,\qquad
		H^z(r) &= B f(r) \,,
	\end{aligned}
 \label{15III19.100}
\end{equation}
where the constants $A$ and $B$ are to be determined from continuity conditions.

The transverse components of the fields are obtained from the $z$ components by
\ptcheck{1III19, together}
\begin{equation}
	\begin{aligned}
		i \zeta E^r 		&= \beta \p_r E^z + \omega \p_\theta H^z/r,\\
		i \zeta E^\theta	&= \beta \p_\theta E^z/r - \omega \p_r H^z,\\
		i \zeta H^r 		&= \beta \p_r H^z - \omega n^2  \p_\theta E^z/r,\\
		i \zeta H^\theta	&= \beta \p_\theta H^z/r + \omega n^2  \p_r E^z,\\
	\end{aligned}
 \label{10IV19.1}
\end{equation}
where we have set
\begin{equation}
\label{22V19.1}
	\zeta = \omega^2 n^2 - \beta^2
	\,,
\end{equation}
Together with the relations \eqref{eq:macroscopic fields unperturbed}, these equations constitute the map $F(f; \beta, \omega)$, which was introduced in the previous section.
Due to the special form of \eqref{eq:macroscopic fields unperturbed}, the continuous fields $E^z$ and $H^z$ satisfy the same wave equation as the fundamental fields $D^z$ and $B^z$. Thus, their continuity can be implemented immediately and the nontrivial continuity conditions reduce to two linearly independent constraints, which can be written in the following form \cite[eq.~(4.41)]{Beig_2018}:
\begin{equation}
	\label{eq:unperturbed:continuity}
	\begin{pmatrix}
		i m \beta \left( \frac{1}{U^2} + \frac{1}{W^2} \right) &
		- \omega \left( \frac{1}{U} \frac{J'_m(U)}{J_m(U)} + \frac{1}{W} \frac{K'_m(W)}{K_m(W)} \right)
		\\
		\omega \left( \frac{n_1^2}{U} \frac{J'_m(U)}{J_m(U)} + \frac{n_2^2}{W} \frac{K'_m(W)}{K_m(W)} \right) &
		i m \beta \left( \frac{1}{U^2} + \frac{1}{W^2} \right)
	\end{pmatrix}
	\begin{pmatrix}
		A \\
		B
	\end{pmatrix}
	= 0
	\,.
\end{equation}
This is the anticipated matrix formulation of the continuity condition $\disc{F} = 0$.

The requirement of \eqref{eq:unperturbed:continuity} to admit nontrivial solutions for $A$ and $B$ leads to the dispersion relation
\begin{equation}
	m^2 \frac{\beta^2}{\omega^2} \left( \frac{1}{U^2} + \frac{1}{W^2} \right)^2
	=
	\left( \frac{1}{U} \frac{J'_m(U)}{J_m(U)} + \frac{1}{W} \frac{K'_m(W)}{K_m(W)} \right)
	\left( \frac{n_1^2}{U} \frac{J'_m(U)}{J_m(U)} + \frac{n_2^2}{W} \frac{K'_m(W)}{K_m(W)} \right)
	\,,
\end{equation}
which must be solved numerically.

\subsubsection{Generalisation}
\tbm{Added on 16V19}
In light of the general formalism of the previous section, it will prove useful to generalise the expressions found above.
To allow for discontinuous fundamental fields $D^z$ and $B^z$, we define
\begin{equation}
	\label{eq: homogeneous solution 0}
	f_0(\alpha_1, \alpha_2; r) =
	\begin{cases}
		\alpha_1 J_m(U r/a),	&	r < a \,,\\
		\alpha_2 K_m(W r/a),	&	r > a \,,
	\end{cases}
\end{equation}
such that the solutions can be written as
\begin{equation}
	\label{eq: unperturbed fundamental fields}
	\begin{aligned}
		D^z(r) &= f_0(\alpha_1, \alpha_2; r),\qquad
		B^r(r) &= f_0(\alpha_3, \alpha_4; r).
	\end{aligned}
\end{equation}
The original expressions are thus obtained by setting
\begin{equation}
	\label{eq: unperturbed parameters}
	\begin{aligned}
		\alpha_1 = \frac{n_1^2 A}{J_m(U)}\,,
		\qquad
		\alpha_2 = \frac{n_2^2 A}{K_m(W)}\,,
		\qquad
		\alpha_3 = \frac{B}{J_m(U)}\,,
		\qquad
		\alpha_4 = \frac{B}{K_m(W)}\,.
	\end{aligned}
\end{equation}
This form of the solution will be used in Chapter~\ref{chapter:first approximation} to determine first order corrections due to Earth’s rotation.
In this calculation, it will prove useful to allow for parameters $\bs \alpha = (\alpha_1, \ldots, \alpha_4)$, which deviate from the unperturbed values given here. 

\section{Main Corrections}
\label{chapter:first approximation}
In this section, we compute the first order corrections to the fields which arise from the approximate wave equation \eqref{eq:wave:approximate} and the modified relation between $\vec e, \vec h$ and $\vec d, \vec b$ given in \eqref{eq:rotation:matter fields}.

To describe corrections to the unperturbed solution, we write
\begin{equation}
	\label{eq:perturbation nomenclature}
	\begin{aligned}
		\vec e &= \vec E + \delta \vec e,\qquad
		\vec d &= \vec D + \delta \vec d,\qquad
		\vec b &= \vec B + \delta \vec b,\qquad
		\vec h &= \vec H + \delta \vec h,
	\end{aligned}
\end{equation}
where $\vec d, \vec b$ etc.\ were defined in \eqref{eq:field strength decomposition}, $\vec D, \vec B$ etc.\ are the “unperturbed” fields given by \eqref{eq: unperturbed fundamental fields} and $\delta \vec d, \delta \vec b$ etc.\ are the first corrections of order $\omlab$.
Note that $\vec D, \vec H$ are related to $\vec E, \vec B$ via $\vec D = n^2 \vec E$ and $\vec B = \vec H$, while $\vec e, \vec h$ are related to $\vec d, \vec b$ via the equations \eqref{eq:rotation:matter fields}.

\subsection{The Wave Equation for the $z$ Component}
In this section, we specialise the vectorial equations \eqref{eq:wave:approximate} to scalar wave equations for the $z$ components of the fields $\vec d$ and $\vec b$.

From now on, we will use cylindrical coordinates $(r, \theta, z)$ and refer all tensorial objects to the orthonormal frame
\begin{equation}
	\label{eq:cylindrical frame}
	\frame_r = \frac{\p}{\p r}\,,
	\qquad
	\frame_\theta = \frac{1}{r} \frac{\p}{\p \theta}\,,
	\qquad
	\frame_z = \frac{\p}{\p z}\,,
\end{equation}
whose coframe is given by
\begin{equation}
	\label{eq:cylindrical coframe}
	\frame^r = \dd r\,,
	\qquad
	\frame^\theta = r \dd \theta\,,
	\qquad
	\frame^z = \dd z\,.
\end{equation}
Since the wave equation \eqref{eq:wave:approximate} is independent of the spatial coordinate system, we may evaluate it in the coordinate system at hand.
As stated in the outline, we are mainly interested in the wave equation for the $z$ components of the fields for which the equation simplifies since $(\Delta \vec b)^z = \Delta (b^z)$ and $(\nabla_i \vec b)^z = \p_i b^z$.
We thus obtain the scalar equations
\begin{equation}
	\label{eq:wave:scalar z equation}
	\begin{aligned}
		n^2 \ddot b^z - \Delta b^z
		&= 2 V(\dot b^z)
		\,,
		\\
		n^2 \ddot d^z - \Delta d^z
		&= 2 V(\dot d^z)
		\,,
	\end{aligned}
\end{equation}
where $V(b^z)$ denotes the action of the vector field $\vec V$ on the scalar function $b^z$ (i.e.~the directional derivative).
This is nothing but the scalar wave equation with respect to the optical metric (using the approximation $\vec v \approx \vec V$).

Inserting the perturbative expansion \eqref{eq:perturbation nomenclature}, neglecting all terms of order $\order(\Omega^2)$, and using the fact that $D^z$ and $B^z$ satisfy the homogeneous wave equation, we obtain
\begin{equation}
	\begin{aligned}
		n^2 \delta \ddot d^z - \Delta \delta d^z
		&= 2 V(\dot D^z)
		\,, \\
		n^2 \delta \ddot b^z - \Delta \delta b^z
		&= 2 V(\dot B^z)
		\,.
	\end{aligned}
\end{equation}

\subsection{The Radial Equations}
\label{section: radial equation}

Summarising, we need to analyse the following differential equations for the $z$ components of $\delta \vec d$ and $\delta \vec b$:
\begin{equation}
	\label{eq:wave equations}
	\begin{aligned}
		n^2 \delta \ddot d^z - \Delta \delta d^z
		&= 2 i \omega V(D^z)
		\,, \\
		n^2 \delta \ddot b^z - \Delta \delta b^z
		&= 2 i \omega V(B^z)
		\,,
	\end{aligned}
\end{equation}
where the fields $D^z$ and $B^z$ are given by
\begin{equation}
	\begin{aligned}
		D^z(t, r, \theta, z) = f_0(\alpha_1, \alpha_2; r) e^{i(\omega t - \beta z + m \theta)}\,,\\
		B^z(t, r, \theta, z) = f_0(\alpha_3, \alpha_4; r) e^{i(\omega t - \beta z + m \theta)}\,.
	\end{aligned}
\end{equation}
Here, the parameter $\beta$ and the coefficients $\alpha_1, \ldots, \alpha_4$ \emph{do not necessarily} coincide with the unperturbed parameters, which would have been obtained when $\Omega = 0$.
Note that in doing so, one modifies the right hand side of the wave equations by terms of order $\order(\Omega^2)$, which we assume to be negligible.
Note also that $V$ on the right hand side depends on the angle $\theta$ via
\tbm{Corrected on 28X19, rest OK}
\begin{equation}
	\begin{aligned}
		V^r			&= V^x \cos \theta + V^y \sin \theta\,,\\
		V^\theta		&= V^y \cos \theta - V^x \sin \theta\,,\\
	\end{aligned}
\end{equation}
where $V^x, V^y$ (and $V^z$) are constants.
Thus, the $\theta$-dependence of the source term is not given by $e^{i m \theta}$ alone; instead, there are \emph{sidebands} with an angular dependence of the form $e^{i (m \pm 1) \theta}$.

Defining the constants $V^\pm$ by
\begin{equation}
	2 V^\pm = i V^x \pm V^y
\end{equation}
and decomposing all fields as a Fourier series in $\theta$ of the form
\begin{equation}
	\label{eq: Fourier series}
	X = \sum_{n = -1}^{+1} \braket{X}_n e^{i (m+n) \theta},
\end{equation}
one has e.g.\
\begin{equation}
	\label{eq:Fourier V decomposition}
	\braket{V^r}_\pm = -i V^\pm \,,
	\qquad
	\braket{V^\theta}_\pm = \pm V^\pm \,,
\end{equation}
so one obtains for $w^i = D^z$ or $w^i = B^z$
\begin{equation}
	\begin{aligned}
		\braket{V(w^i)}_0 &= - i \beta V^z w^i\,,\\
		\braket{V(w^i)}_\pm &= - i V^\pm c_m^\pm w^i\,,
	\end{aligned}
\end{equation}
where we have set
\begin{equation}
	\label{eq:operators pm}
	c_m^\pm = \p_r \mp \frac{m}{r}\,.
\end{equation}
For the perturbed fields $\delta d^z, \delta b^z$ we use the ansatz
\begin{equation}
	\label{eq: Fourier ansatz}
	\begin{aligned}
		\delta d^z(t, r, \theta, z) &= \left[
		\delta d^z_0(r) + \delta d^z_+(r) e^{+ i \theta} + \delta d^z_-(r) e^{- i \theta}
		\right]e^{i(\omega t - \beta z + m \theta)}\,,\\
		\delta b^z(t, r, \theta, z) &= \left[
		\delta b^z_0(r) + \delta b^z_+(r) e^{+ i \theta} + \delta b^z_-(r) e^{- i \theta}
		\right]e^{i(\omega t - \beta z + m \theta)}\,,
	\end{aligned}
\end{equation}
which leads to the following set of equations for the radial functions:
\begin{equation}
	\label{02VI19.4}
	\begin{aligned}
		\bessel{\zeta}{m} \delta b^z_0(r) &= -2 \beta \omega V^z B^z(r),\\
		\bessel{\zeta}{m} \delta d^z_0(r) &= -2 \beta \omega V^z D^z(r),\\
		\bessel{\zeta}{m \pm 1} \delta b^z_\pm(r) &= -2 \omega V^\pm c_m^\pm B^z(r),\\
		\bessel{\zeta}{m \pm 1} \delta d^z_\pm(r) &= -2 \omega V^\pm c_m^\pm D^z(r),\\
	\end{aligned}
\end{equation}
where $\zeta = n^2 \omega^2 - \beta^2$ as before and
\begin{equation}
	\bessel{\zeta}{\nu} f(r) :=
	f''(r) + \frac{1}{r} f'(r) + \left( \zeta - \frac{\nu^2}{r^2} \right) f(r)
\end{equation}
is either an ordinary or modified Bessel operator, depending on whether $\zeta > 0$ or $\zeta < 0$.
We consider only the equations for $\vec d$, since those for $\vec b$ are identical. Using \eqref{eq: unperturbed fundamental fields}, we obtain equations of the form
\begin{equation}
	\label{eq:differential eq:p0, ppm}
	\begin{aligned}
		\bessel{\zeta}{m} p_0(\alpha_1, \alpha_2; r)
		&= a^{-2} f_0(\alpha_1, \alpha_2; r)
		\,,\\
		\bessel{\zeta}{m\pm1} p_\pm(\alpha_1, \alpha_2; r)
		&= a^{-1} c_m^\pm f_0(\alpha_1, \alpha_2; r)
		\,,
	\end{aligned}
\end{equation}
where the coefficients in front of the source terms have been dropped momentarily – they can be added later due to the linearity of the equations.
The powers of $a$ were chosen such that the functions $p_0(\alpha_i, \alpha_j; r)$ and $p_\pm(\alpha_i, \alpha_j; r)$ have the same dimension as $f_0(\alpha_i, \alpha_j; r)$, which, in turn, has the same dimension as the coefficients $\alpha_i, \alpha_j$.
The homogeneous solutions of interest (we require the functions to be finite at the origin and to decay sufficiently fast at large distances) to the first equation are given by \eqref{eq: homogeneous solution 0}. The corresponding homogeneous solutions for the sidebands are then given by
\begin{equation}
	\label{eq: homogeneous solution pm}
	f_\pm(\chi_1, \chi_2; r) =
	\begin{cases}
		\mp U \chi_1 J_{m \pm 1}(U r/a),& r < a \,,\\
		- W \chi_2 K_{m \pm 1}(W r/a),& r > a \,.\\
	\end{cases}
\end{equation}
The factors $\mp U$ and $-W$ were chosen such that
\begin{equation}
	f_\pm(\alpha_i, \alpha_j; r)
	= a c_m^\pm f_0(\alpha_i, \alpha_j; r) \,,
\end{equation}
which is due to the following recursion relations
\begin{equation}
	\label{eq:Bessel recursion}
	\begin{aligned}
		J_m'(r) \mp \frac{m}{r} J_m(r) = \mp J_{m \pm 1}(r)\,,
		\qquad
		K_m'(r) \mp \frac{m}{r} K_m(r) = - K_{m \pm 1}(r)\,,
	\end{aligned}
\end{equation}
cf.~\cite[p.~361, Equation~9.1.27]{Abramowitz_1964} and \cite[p.~376, Equation~9.6.26]{Abramowitz_1964}.

\begin{figure}[h]
	\centering
	\subfloat[][Typical behaviour in the core ($r < a$)]{\includegraphics[width=0.48\columnwidth]{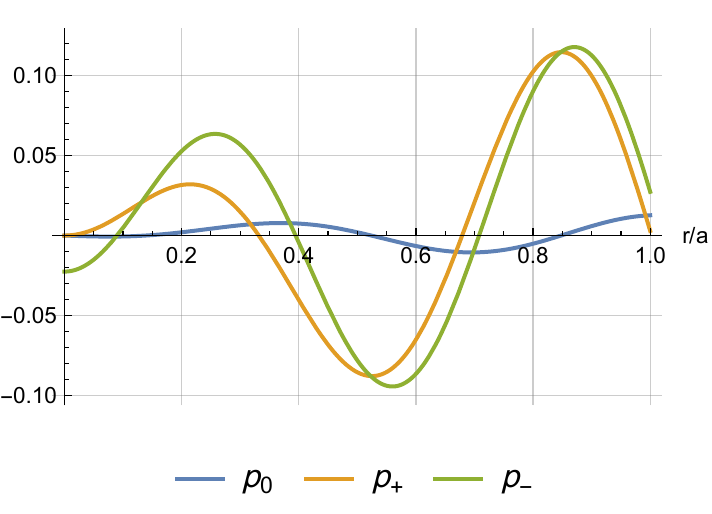}}
	\hfill
	\subfloat[][Typical behaviour in the cladding ($r > a$)]{\includegraphics[width=0.48\columnwidth]{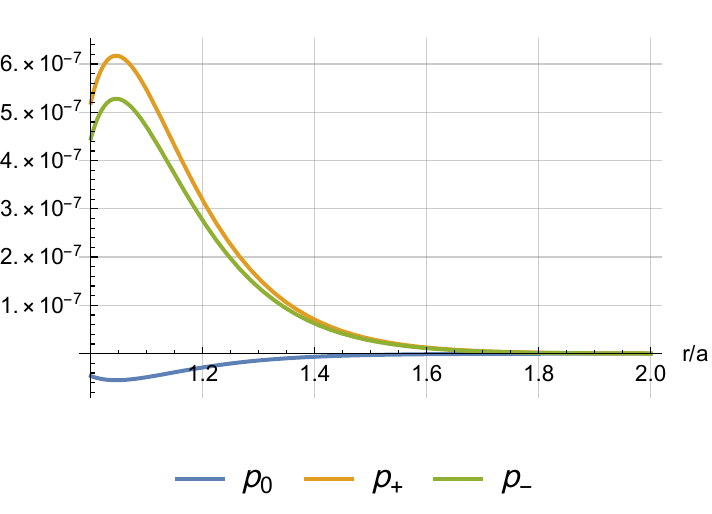}}
	\caption{Plots of the functions $p_0$ and $p_\pm$ for $m = 1$ and $U = W = 10$.}
	\label{plot:particular solutions p}
\end{figure}

Particular solutions $p_0$ and $p_\pm$ which also decay at large distances and do not blow up at $r = 0$ are given by
\begin{equation}
	\label{eq: particular solutions}
	\begin{aligned}
		p_0(\alpha_1, \alpha_2; r)
		&=
		\begin{cases}
		+ \alpha_1 \frac{\pi}{2}
		[J_m(U r/a) \Gamma_0(r/a) + Y_m(U r/a) \Delta_0(r/a)]
			&	r < a,\\
		- \alpha_2
		[ K_m(W r/a) \Sigma_0(r/a) + I_m(W r/a) T_0(r/a) ]
			&	r > a,
		\end{cases}\\
		p_+(\alpha_1, \alpha_2; r)
		&=
		\begin{cases}
		- \alpha_1 U \frac{\pi}{2}
		[J_{m+1}(U r/a) \Gamma_+(r/a) + Y_{m+1}(U r/a) \Delta_+(r/a)]
			&	r < a,\\
		+ \alpha_2 W
		[ K_{m+1}(W r/a) \Sigma_+(r/a) + I_{m+1}(W r/a) T_+(r/a) ]
			&	r > a,\\
		\end{cases}\\
		p_-(\alpha_1, \alpha_2; r)
		&=
		\begin{cases}
		+ \alpha_1 U \frac{\pi}{2}
		[ J_{m-1}(U r/a) \Gamma_-(r/a) + Y_{m-1}(U r/a) \Delta_-(r/a)]
			&	r < a,\\
		+ \alpha_2 W
		[ K_{m-1}(W r/a) \Sigma_-(r/a) + I_{m-1}(W r/a) T_-(r/a) ]
			&	r > a,
		\end{cases}
	\end{aligned}
\end{equation}
where we have introduced the functions
\begin{equation}
	\label{eq: def Gamma Delta Sigma Tau}
	\begin{aligned}
		\Gamma_n(z) &= \int_z^1 \xi\, Y_{m+n}(U \xi) J_{m+n}(U \xi) \dd \xi\,,\\
		\Delta_n(z) &= \int_0^z \xi J_{m+n}(U \xi)^2 \dd \xi\,,\\
		\Sigma_n(z) &= \int_1^z \xi I_{m+n}(W \xi) K_{m+n}(W \xi) \dd \xi,\\
		T_n(z) &= \int_z^\infty \xi K_{m+n}(W \xi)^2 \dd \xi\,,
	\end{aligned}
\end{equation}
which can be evaluated explicitly using the indefinite integrals (up to additive constants)
\begin{equation}
	\begin{aligned}
		\int J_m(r)^2 \,r\dd r
		&= \frac{r^2}{2} \left( J_m(r)^2 - J_{m-1}(r) J_{m+1}(r) \right)\,,\\
		\int K_m(r)^2 r\dd  r
		&= \frac{r^2}{2} \left( K_m(r)^2 - K_{m-1}(r) K_{m+1}(r) \right)\,,\\
		\int Y_m(r) J_m(r) \,r\dd r
		&= \frac{r^2}{2} \left( J_m(r) Y_m(r) - J_{m-1}(r) Y_{m+1}(r) \right)\,,\\
		\int I_m(r) K_m(r) \,r\dd r
		&= \frac{r^2}{2} \left( K_m(r) I_m(r) + K_{m-1}(r) I_{m+1}(r) \right)\,.
	\end{aligned}
\end{equation}
The solution to the equations \eqref{02VI19.4} are thus
\begin{equation}
	\label{eq:perturbed fundamental fields result}
	\begin{aligned}
		\delta d^z_0(r)
		&=
			- 2 a^2 \beta \omega V^z p_0(\alpha_1, \alpha_2; r)
		\,,\\
		\delta b^z_0(r)
		&=
			- 2 a^2 \beta \omega V^z p_0(\alpha_3, \alpha_4; r)
		\,,\\
		\delta d^z_\pm(r)
		&=
			- 2 a \omega V^\pm p_\pm(\alpha_1, \alpha_2; r)
			+ f_\pm(\chi_1^\pm, \chi_2^\pm; r)
		\,,\\
		\delta b^z_\pm(r)
		&=
			- 2 a \omega V^\pm p_\pm(\alpha_3, \alpha_4; r)
			+ f_\pm(\chi_3^\pm, \chi_4^\pm; r)
		\,.
	\end{aligned}
\end{equation}
The functions $p_0$ and $p_\pm$ are plotted in figure~\ref{plot:particular solutions p}.
As expected, the functions are regular at the origin and decay rapidly in the cladding.

\subsection{Compatibility of Approximations}
\label{section: compatibility}

Until now we were only concerned with the $z$ components of the fields. In all calculations we have omitted terms quadratic in $\Omega$ and we have neglected some terms leading to relative errors of the order of $a/R$ or $\ell/R$, where $a$ denotes the radius of the dielectric, $\ell$ its length and $R$ denotes Earth's radius.

In the next section we will compute the remaining components using similar approximations.
There is however a subtlety: as we will now show, these approximations cannot be discussed term by term, since the resulting expressions must satisfy two equations, which relate three field components each.

We shall later require the continuity of six field components at $r = a$ (the core-cladding interface) due to the following conditions:
\begin{enumerate}
  \item The non-existence of magnetic surface charges and surface currents is equivalent to the continuity of $  b^r $, $e^\theta$ and $e^z$; and
  \item The absence of electric surface charges and surface currents is  equivalent to the continuity of $d^r$, $h^\theta$ and $ h^z$.
\end{enumerate}

Clearly, the continuity conditions must hold for all Fourier modes separately.
Note, however, that these conditions constitute six linear equations in the four coefficients which parameterise the solutions in each mode.
In order to obtain nontrivial solutions, two continuity conditions must therefore be expressible in terms of the remaining four equations.
The existence of such a relation follows in fact from the $r$-components of \eqref{eq:maxwell},
\begin{align}
	\label{eq:consistency conditions1}
	i \omega b^r + \frac{1}{r} \p_\theta e^z - \p_z e^\theta  &=  0
	\,,
	\\
	\label{eq:consistency conditions2}
	i \omega d^r - \frac{1}{r} \p_\theta h^z + \p_z h^\theta   &=  0
	\,.
\end{align}
One sees that, for solutions as considered in this work, if these two equations hold everywhere, and if the $b^r, d^r, h^z, e^z$ are continuous, then $h^\theta$ and $e^\theta$ will also be continuous.
Note that continuity of $b^r, d^r, h^\theta, e^\theta$ does not imply continuity of $h^z$ and $e^z$, since they may be independent of $\theta$.

In order to obtain consistent equations, it is therefore necessary to make approximations which are compatible with these two equations.

The components $e^z, h^z$ are determined by \eqref{eq:rotation:matter fields} whereas the components $b^r, d^r, h^\theta, e^\theta$ will be computed from \eqref{eq:rotation:maxwell covariant}.
The approximations made in either set of equations thus determines the approximations in the other one.

\subsection{The Transverse Components of $d$ and $b$}

\label{section: transverse components D, B}
We use the equations \eqref{eq:maxwell} to express the “transverse” $r$ and $\theta$ components of the fields in terms of the “longitudinal” $z$ component.
Substituting \eqref{eq:perturbation nomenclature}, one obtains to first order in $\omlab$
\begin{equation}	
	\label{24V19.1}
	\begin{aligned}
		n^2 \delta \dot{\vec b}
		+ \nabla \times \delta \vec d 
		- [\vec v, \vec H] &= 0
		\,,\\
		\delta \dot{\vec d}
		- \nabla \times \delta \vec b
		- [\vec v, \vec E] &= 0
		\,,
	\end{aligned}
\end{equation}
where we have used $\mu = 1$, $\varepsilon = n^2$, $\vec D = n^2 \vec E$, $\vec B = \vec H$ and fact that $n^2$ is locally constant.
If all field components depend on $t$ and $z$ via $\exp(i(\omega t - \beta z))$, the differential equations for the transverse components reduce to the following set of linear algebraic equations
\begin{equation}
	\begin{aligned}
		i n^2 \omega \delta b^r + i \beta \delta d^\theta &= [\vec v, \vec H]^r - \p_\theta \delta d^z/r,\\
		i n^2 \omega \delta b^\theta- i \beta \delta d^r &= [\vec v, \vec H]^\theta + \p_r \delta d^z,\\
		i \omega \delta d^r - i \beta \delta b^\theta &=[\vec v, \vec E]^r + \p_\theta \delta b^z/r,\\
		i \omega \delta d^\theta + i \beta \delta b^r &= [\vec v, \vec E]^\theta - \p_r \delta b^z,
	\end{aligned}
\end{equation}
with the solution
\begin{equation}
	\label{12IV19.1}
	\begin{aligned}
		i \zeta \delta d^r
		&= n^2 \omega ([\vec v, \vec E]^r + \p_\theta \delta b^z/r)
		+ \beta ([\vec v, \vec H]^\theta + \p_r \delta d^z) \,,\\
		i \zeta \delta d^\theta
		&= n^2 \omega ([\vec v, \vec E]^\theta - \p_r \delta b^z)
		- \beta ([\vec v, \vec H]^r - \p_\theta \delta d^z/r) \,,\\
		i \zeta \delta b^r
		&= \omega ([\vec v, \vec H]^r - \p_\theta \delta d^z/r)
		- \beta ([\vec v, \vec E]^\theta - \p_r \delta b^z) \,,\\
		i \zeta \delta b^\theta
		&= \omega ([\vec v, \vec H]^\theta + \p_r \delta d^z)
		+ \beta ([\vec v, \vec E]^r + \p_\theta \delta b^z/r) \,,
	\end{aligned}
\end{equation}
where $\zeta = n^2 \omega^2 - \beta^2$.
Note that for any vector field $\vec w$ it holds that
\begin{equation}
	[\vec v, \vec w] = \Ds_{\vec v}\, \vec w - \vecomlab \times \vec w\,,
\end{equation}
where $\Ds$ denotes the Levi-Civita covariant derivative with respect to the spatial metric.
\ptcheck{ 10IV19 up to here in this section}

As a first approximation, we write
\begin{equation}
	[\vec v, \vec w] \approx \nabla_{\vec V}\, \vec w \,.
\end{equation}
The approximation $\nabla_{\vec v}\, \vec w \approx \nabla_{\vec V}\, \vec w$ corresponds to $\vec x \approx \vec R$, where relative errors are expected to be of order $a/R$ or $\ell/R$.
Moreover, neglecting $\vec \Omega \times \vec w$ compared to the first term is expected to produce relative errors of the order of $\lambda/R$, cf.\ the discussion in Section~\ref{s:EM wave equation}.

Defining the connection one-forms
\begin{equation}
	\t\omega{^a_b}(X)
	= \frame^a(\Ds_X \frame_b) \,,
\end{equation}
we may decompose the covariant derivative as
\begin{equation}
	(\Ds_X Y)^a = X(Y^a) + \t\omega{^a_b}(X) Y^b
	\,.
\end{equation}
We find the only non-zero connection forms to be
\begin{equation}
	\t\omega{^\theta_r} = + \frac{1}{r} \frame^\theta \,,
	\qquad
	\t \omega{^r_\theta} = - \frac{1}{r} \frame^\theta \,.
\end{equation}
Since the first order perturbation fields have three $\theta$ modes, e.g.
\begin{equation}
	\delta d^z = e^{i(\omega t - \beta z + m \theta)} \sum_{n = -1}^{+1} \delta d_n(r) e^{i n \theta},
\end{equation}
we decompose all expressions in a Fourier series in $\theta$ using the notation introduced in \eqref{eq: Fourier series}. As before, we use the abbreviation $\braket{X}_\pm = \braket{X}_{\pm 1}$.
Note that the cylindrical frame components of $E^\theta$, $E^r$, $B^\theta$ and  $B^r$ of the unperturbed fields have vanishing $m\pm 1$ components by \eqref{10IV19.1}.

Using the notation \eqref{eq:Fourier V decomposition} (and a similar notation for $\vec \Omega$), we obtain the following decomposition of the Lie bracket $[\vec w, \vec v]$, where $\vec w$ is any unperturbed field:
\tbmr{$\pm$ modes need rigorous checking \\ -- \\ ptc: done on 10IV}
\tbm{Modified on 8V19: Removed $\Omega^z$ terms}
\begin{equation}
	\label{12IV19.2}
	\begin{aligned}
		\braket{[\vec w,\vec v]^r}_0\,
			&= i \beta V^z w^r ,\\
		\braket{[\vec w,\vec v]^\theta}_0\,
			&= i \beta V^z w^\theta ,\\
		\braket{[\vec w,\vec v]^z}_0\,
			&= i \beta V^z w^z,\\
		\braket{[\vec w,\vec v]^r}_\pm
			&= i V^\pm c_m^\pm w^r \pm V^\pm w^\theta/r \,,\\
		\braket{[\vec w,\vec v]^\theta}_\pm
			&= i V^\pm c_m^\pm w^\theta \mp V^\pm w^r/r \,,
	\end{aligned}
\end{equation}
where we have used the operator $c_m^\pm = \p_r \mp m/r$, defined in \eqref{eq:operators pm}.

Combining \eqref{12IV19.1} and \eqref{12IV19.2}, one obtains the following expressions for the $m$ modes of the electric field $\vec d$ and the magnetic field $\vec h$
\tbm{Modified on 8V19: Removed $\Omega^z$ terms}
\begin{equation}
	\label{eq: perturbed components A (m mode)}
	\begin{aligned}
		i \zeta \delta d^r_0
		&=
		\beta \p_r \delta d^z_0 + i \omega \frac{m}{r} n^2 \delta b^z_0
		- i V^z \beta ( \beta H^\theta + \omega D^r ) \,,\\
		i \zeta \delta d^\theta_0
		&=
		i \beta \frac{m}{r}  \delta d^z_0 - \omega n^2 \p_r \delta b^z_0
		+ i V^z \beta ( \beta H^r - \omega D^\theta ) \,,\\
		i \zeta \delta b^r_0
		&= \beta \p_r \delta b^z_0 - i \omega \frac{m}{r} \delta d^z_0
		+ i V^z \beta ( \beta E^\theta - \omega H^r ) \,,\\
		i \zeta \delta b^\theta_0
		&=
		i \beta \frac{m}{r} \delta b^z_0 + \omega \p_r \delta d^z_0
		- i V^z \beta ( \beta E^r + \omega H^\theta ) \,,
	\end{aligned}
\end{equation}
 \ptcheck{24IV19 together}
and similarly for the $m \pm 1$ modes
\begin{equation}
	\label{eq: perturbed components A (sidebands)}
	\begin{aligned}
		i \zeta \delta d^r_\pm
		&=
		\beta \p_r \delta d^z_\pm + i \omega \frac{m\pm 1}{r} n^2 \delta b^z_\pm
		- i V^\pm c_m^\pm( \beta H^\theta + \omega D^r )
		\pm \frac{V^\pm}{r} ( \beta H^r - \omega D^\theta ) \,,\\
		i \zeta \delta d^\theta_\pm
		&=
		i \beta \frac{m\pm1}{r} \delta d^z_\pm - \omega n^2 \p_r \delta b^z_\pm
		+ i V^\pm c_m^\pm( \beta H^r - \omega D^\theta )
		\pm \frac{V^\pm}{r} ( \beta H^\theta + \omega D^r) \,,\\
		i \zeta \delta b^r_\pm
		&=
		\beta \p_r \delta b^z_\pm - i \omega \frac{m\pm 1}{r} \delta d^z_\pm
		+ i V^\pm c_m^\pm( \beta E^\theta - \omega H^r )
		\mp \frac{V^\pm}{r} ( \beta E^r + \omega H^\theta ) \,,\\
		i \zeta \delta b^\theta_\pm
		&=
		i \beta \frac{m\pm 1}{r} \delta b^z_\pm + \omega \p_r \delta d^z_\pm
		- i V^\pm c_m^\pm( \beta E^r + \omega H^\theta )
		\mp \frac{V^\pm}{r} ( \beta E^\theta - \omega H^r ) \,.
	\end{aligned}
\end{equation}
 \ptcheck{24IV19 together}

\subsection{The Field Components of $e$ and $h$}

We rewrite equation \eqref{eq:rotation:matter fields} with $\mu =1$ and $\varepsilon = n^2$ in the form
\begin{equation}
	\begin{aligned}
		n^2 \vec e
		&= \vec d
		+ \vec v \times \vec b \,,
		\\
		n^2 \vec h
			&= n^2\vec b
			- \vec v \times \vec d \,,
	\end{aligned}
\end{equation}
where $\vec v = \vecomlab \times \vec x$.
Substituting the perturbative expansion \eqref{eq:perturbation nomenclature} and using $\vec D = n^2 \vec E$ as well as $\vec B = \vec H$, one obtains to first order in $\omlab$
\begin{equation}
	\begin{aligned}
		n^2 \delta \vec e
		&= \delta \vec d
		+ \vec v \times \vec B \,,
		\\
		\delta \vec h
		&= \delta \vec b
		- \vec v \times \vec E \,.
	\end{aligned}
 \label{23V19.21}
\end{equation}
We compute the $\theta$ and $z$ components of these fields, which are required to be continuous.
Using the approximation $\vec v \approx \vec V$ (in accordance with $\vec x = \vec R + \vec r \approx \vec R)$, one finds for the $m$ modes
\begin{equation}
	\label{eq: perturbed components B (m mode)}
	\begin{aligned}
		n^2 \delta e^\theta_0
			&= \delta d^\theta_0 + V^z B^r \,,\\
		n^2 \delta e^z_0
			&= \delta d^z_0 \,,\\
		\delta h^\theta_0
			&= \delta b^\theta_0 - V^z E^r \,,\\
		\delta h^z_0
			&= \delta b^z_0 \,,
	\end{aligned}
\end{equation}
and for the $m \pm 1$ modes
\begin{equation}
	\label{eq: perturbed components B (sidebands)}
	\begin{aligned}
		n^2 \delta e^\theta_\pm
			&= \delta d^\theta_\pm + i V^\pm B^z \,,\\
		n^2 \delta e^z_\pm
			&= \delta d^z_\pm - i V^\pm (B^\theta \mp i B^r) \,,\\
		\delta h^\theta_\pm
			&= \delta b^\theta_\pm - i V^\pm E^z \,,\\
		\delta h^z_\pm
			&= \delta b^z_\pm + i V^\pm ( E^\theta \mp i E^r) \,.
	\end{aligned}
\end{equation}
\ptcheck{24IV19 together}

\subsection{Consistency of the Approximations}
\label{s:consistency 1}
\ptcheck{23V19 the whole section}

In this section, we verify that the approximations made are consistent with the equations \eqref{eq:consistency conditions1}-\eqref{eq:consistency conditions2}.
Due to the electromagnetic symmetry, it suffices to verify the equation
\begin{equation}
	\label{23V19.11}
	i \omega d^r
	- \frac{1}{r} \p_\theta h^z
	+ \p_z h^\theta = 0
	\,.
\end{equation}

Let us start by showing that if all fields depend on $z$ only through $e^{-i \beta z}$, and if the field components are related to each other as in the unperturbed case, i.e.\ the $r$ and $\theta$ components are related to the $z$ components by \eqref{10IV19.1} (or equivalently by \eqref{12IV19.1} without the Lie bracket terms) and by $\vec h = \vec b$, $\vec d = n^2 \vec e$, then  \eqref{eq:consistency conditions1}-\eqref{eq:consistency conditions2} hold.

Hence, we  consider the following subset of the exact Maxwell’s equations with $\Omega\equiv 0$:
\begin{equation}
	\begin{aligned}
		i \zeta D^r
		&= n^2 \omega \frac{1}{r} \p_\theta B^z + \beta \p_r D^z\,,\\
		i \zeta B^\theta
		&= \omega \p_r D^z + \frac{1}{r} \beta \p_\theta B^z\,,\\
		H^\theta &= B^\theta\,,\\
		H^z &= B^z
 \,,
	\end{aligned}
 \label{23V19.12}
\end{equation}
with $\zeta$  defined in \eqref{22V19.1},
where all fields depend on $z$ only as $e^{-i \beta z}$.
We have
\begin{equation}
	\begin{aligned}
		i \zeta ( i \omega D^r + \p_z H^\theta)
		&=
		i \zeta ( i \omega D^r -i \beta H^\theta)\\
		&=
		\frac{i}{r} n^2 \omega^2 \p_\theta B^z + i\beta \omega \p_r D^z
		- i\beta \omega \p_r D^z
		- \frac{i}{r} \beta^2 \p_\theta B^z
		= i \zeta \frac{1}{r} \p_\theta B^z
		\,.
	\end{aligned}
 \label{23V19.13}
\end{equation}
Equation \eqref{23V19.11} with the $\vec d$ and $\vec h$ fields replaced by $\vec D$ and $\vec H$ now follows from the last equation in \eqref{23V19.12}, namely $H^z = B^z$.

In fact, the calculation in \eqref{23V19.13}   shows  that if a set of equations with a structure as in \eqref{23V19.12} holds, then \eqref{23V19.11} will hold  \emph{regardless} of the precise form of the dependence upon $r$ and $\theta$ of the $z$ components of the fields.
 
Let us return to the problem at hand. As such, \eqref{12IV19.1} can be written as
\begin{equation}
	\begin{aligned}
		i \zeta d^r
		&= n^2 \omega \frac{1}{r} \p_\theta b^z + \beta \p_r d^z + ... \,,\\
		i \zeta b^\theta
		&= \omega \p_r d^z + \frac{1}{r} \beta \p_\theta b^z + .. \,, 
	\end{aligned}
 \label{23V19.14}
\end{equation}
while \eqref{23V19.21} reads 
\begin{equation}
	\begin{aligned} 
		h^\theta &= b^\theta + ... \,,\\
		h^z &= b^z + ... 
 \,,
	\end{aligned}
 \label{23V19.15}
\end{equation}
The calculation in \eqref{23V19.13} applies to those terms in \eqref{23V19.14}-\eqref{23V19.15} which have been written out explicitly.
Hence, to verify \eqref{23V19.11} it remains to check that an identity of the form \eqref{23V19.11} is satisfied by the terms, denoted by $...$ in \eqref{23V19.14}-\eqref{23V19.15}, which have not been written in explicit form.

Equivalently, it remains to consider the deviations from the “unperturbed relations” when checking \eqref{23V19.11}.
Let us use the notation $x \leadsto y$ when $x$ (e.g.\ $\delta d^r$) deviates from the “unperturbed expression”  by $y$.
For instance, for the $m$ mode we have by equations \eqref{eq: perturbed components A (m mode)} and \eqref{eq: perturbed components B (m mode)}, suppressing the common factor $- V^z$ temporarily
\begin{equation}
	\begin{aligned}
		i \zeta \delta d^r_0
		&\leadsto i \beta (\beta H^\theta + \omega D^r)\,,\\
		i \zeta \delta b^\theta_0
		&\leadsto i \beta (\beta E^r + \omega H^\theta)\,,\\
		\p_z \delta h^\theta_0 + i \beta \delta b^\theta_0
		&\leadsto \p_z E^r\,,\\
		\delta h^z_0
		&\leadsto 0
 \,.
	\end{aligned}
\end{equation}
Since $\zeta = n^2 \omega^2 - \beta^2$ it follows that
\begin{equation}
	i \omega \delta d^r_0
	- i \beta \delta b^\theta_0
	\leadsto i \beta E^r\,,
\end{equation}
and thus
\begin{equation}
	i \omega \delta d^r_0 + \p_z \delta h^\theta_0
	= (i \omega \delta d^r_0 - i \beta \delta b^\theta_0) + (\p_z \delta h^\theta_0 - i \beta \delta b^\theta_0)
	\leadsto
	 i \beta E^r + \p_z E^r = 0
	\,,
\end{equation}
which establishes the consistency of the equations for the $m$ mode.

For the sidebands, we have  from \eqref{eq: perturbed components A (sidebands)} and 
\eqref{eq: perturbed components B (sidebands)}, and after omitting the common factor $- V^\pm$,
\begin{equation}
	\begin{aligned}
		i \zeta \delta d^r_\pm 
		&\leadsto
		i c_m^\pm(\beta H^\theta + n^2 \omega E^r)
		\mp \frac{1}{r}(\beta H^r - n^2 \omega E^\theta) \,,
		\\
		i \zeta \delta b^\theta_\pm 
		&\leadsto
		i c_m^\pm( \beta E^r + \omega H^\theta)
		\pm \frac{1}{r}( \beta E^\theta - \omega H^r) \,,
		\\
		\p_z \delta h^\theta_\pm + i \beta \delta b^\theta_\pm
		&\leadsto
		i \p_z E^r \,,
		\\
		\delta h^z_\pm 
		&\leadsto
		-i E^\theta \mp E^r \,.
	\end{aligned}
\end{equation}
Using the definition $c_m^\pm = \p_r \mp m/r$, one obtains
\begin{equation}
	i \omega \delta d^r_\pm  - i \beta \delta b^\theta_\pm 
	\leadsto i c_m^\pm E^r \pm \frac{1}{r} E^\theta
	= i \p_r E^r \mp \frac{i}{r}m E^r \pm \frac{1}{r} E^\theta \,.
\end{equation}
Thus,
\begin{equation}
	\begin{aligned}
		i \omega \delta d^r_\pm
		+ \p_z \delta h^\theta_\pm
		- i\frac{(m \pm 1)}{r} \delta h^z_\pm
		&= (i \omega \delta d^r_\pm - i \beta \delta b^\theta_\pm)
		+ (\p_z \delta h^\theta_\pm + i \beta \delta b^\theta_\pm)
		- i\frac{m \pm 1}{r} \delta h^z_\pm
		\\
		&\leadsto
		i \p_r E^r
		\mp i\frac{m}{r} E^r
		\pm \frac{1}{r} E^\theta
		+ i \p_z E^r
		+  \frac{m \pm 1}{r} (\pm i E^r - E^\theta)
		\\
		&=
		i \left( \p_r E^r + \frac{1}{r}E^r + \frac{1}{r} \p_\theta E^\theta + \p_z E^z \right)
		= i (\vec \nabla \cdot \vec E)
		= 0 \,.
	\end{aligned}
\end{equation}
This completes the proof of  the consistency of the approximate equations for the sidebands.
 \ptcheck{23V19 the whole section}

\subsection{Continuity Conditions}
In this section, we discuss a general method to analyse the continuity conditions of the various field components.

According to the discussion in Section~\ref{section: compatibility}, the requirement that $d^r, e^\theta, e^z$ and $b^r, h^\theta, h^z$ be continuous at the core-cladding interface constitute only \emph{four} linearly independent equations.
It thus suffices to require $d^r, e^z, b^r$ and $h^z$ to be continuous, so we define a vector which summarises the relevant jump heights:
\begin{equation}
	\disc{F}
	=
	\begin{pmatrix}
		\disc{d^r} &
		\disc{e^z} &
		\disc{b^r} &
		\disc{h^z}
	\end{pmatrix}^\mathsf{T}
	\,,
\end{equation}
where $x^\mathsf{T}$ denotes the transpose of $x$.
Here, we have introduced the notation $\disc{f}$ for the jump height of a function $f$ at $r = a$:
\begin{equation}
	\disc{f} := \lim_{r \nearrow a} f(r) - \lim_{r \searrow a} f(r) \,.
\end{equation}

\subsubsection*{Discontinuity Matrices}

Due to linearity of the system, $\disc{F}$ is a linear homogeneous function of the coefficients
\begin{equation}
	\begin{aligned}
		\boldsymbol \alpha
		&= \begin{pmatrix}
			\alpha_1 &
			\alpha_2 &
			\alpha_3 &
			\alpha_4
		\end{pmatrix}^\mathsf{T}
		\,,
		\\
		\boldsymbol \chi^\pm
		&= \begin{pmatrix}
			\chi_1 &
			\chi_2 &
			\chi_3 &
			\chi_4
		\end{pmatrix}^\mathsf{T}
		\,,
	\end{aligned}
\end{equation}
which parameterise the fields according to \eqref{eq: unperturbed fundamental fields} and \eqref{eq:perturbed fundamental fields result}.
Thus, there exist (complex) $4 \times 4$ matrices $M_1, M_2^\pm$ such that
\begin{equation}
	\disc{F} = M_1 \bs \alpha + M_2^\pm \boldsymbol \chi^\pm
	\,.
\end{equation}
We will refer to these matrices as \emph{discontinuity matrices}.
In fact, equations of this form hold for every Fourier mode of the fields separately.

To compute these matrices, we consider the jump height of a field component $w$ as a linear map of the coefficients (e.g.\ $\bs \alpha$) to the field of complex numbers.
We may thus identify $\disc{w}$ with a row vector of four complex entries, such that
\begin{equation}
	\disc{w} = \disc{w}^i \alpha_i
	\,,
\end{equation}
for which we write
\begin{equation}
	\disc{w} \doteq
	\begin{pmatrix}
		\disc{w}^1 & \disc{w}^2 & \disc{w}^3 & \disc{w}^4
	\end{pmatrix}_{\bs\alpha}
	\,.
\end{equation}
The subscript ${\bs\alpha}$ indicates that the form acts on the coefficients $\bs\alpha$, and not on the coefficients $\bs \chi^\pm$.
We will omit this subscript whenever there is no danger of confusion.

Using this notation, the rows of $M_1$ in the above equation are exactly the row vectors which represent $\disc{d^r}, \disc{e^z}, \disc{b^r}$ and $\disc{h^z}$, respectively.
Using the expressions \eqref{eq: homogeneous solution 0} and \eqref{eq: unperturbed fundamental fields}, we find jump heights of the unperturbed fields to be given by
\begin{equation}
	\label{eq:discontinuities unperturbed fields}
	\begin{aligned}
		\disc{D^z}
		&\doteq
		\begin{pmatrix}
			J_m(U) & - K_m(W) & 0 & 0
		\end{pmatrix}
		\,,
		\\
		\disc{B^z}
		&\doteq
		\begin{pmatrix}
			0 & 0 & J_m(U) & - K_m(W)
		\end{pmatrix}
		\,,
		\\
		\disc{\p_r D^z}
		&\doteq
		\begin{pmatrix}
			\frac{U}{a} J_m'(U) & - \frac{W}{a} K_m'(W) & 0 & 0
		\end{pmatrix}
		\,,
		\\
		\disc{\p_r B^z}
		&\doteq
		\begin{pmatrix}
			0 & 0 & \frac{U}{a} J_m'(U) & - \frac{W}{a} K_m'(W)
		\end{pmatrix}
		\,.
	\end{aligned}
\end{equation}
Since the unperturbed fields depend on $\alpha_1, \ldots, \alpha_4$ only, it is clear that the suppressed subscript is $\bs \alpha$.

Multiplication of a field with a constant clearly multiplies the corresponding row vector by the same number.
If a field component $w$, whose jump heights are given by
\begin{equation}
	\disc{w} \doteq
	\begin{pmatrix}
		w^1 & w^2 & w^3 & w^4
	\end{pmatrix}
	\,,
\end{equation}
is multiplied by a function $f$, whose left- and right-sided limits
\begin{equation}
	f(a_-) := \lim_{r \nearrow a} f(r)
	\,,
	\qquad
	f(a_+) := \lim_{r \searrow a} f(r)
\end{equation}
exist, then the jump height of $f w$ is given by
\begin{equation}
	\disc{f w} \doteq
	\begin{pmatrix}
		f(a_-) w^1 &
		f(a_+) w^2 &
		f(a_-) w^3 &
		f(a_+) w^4
	\end{pmatrix}
	\,.
\end{equation}
For example, if $f = n^2$ or $f = \zeta$, then
\begin{equation}
	\begin{aligned}
		\disc{n^2 w}
		&\doteq
		\begin{pmatrix}
			n_1^2 w^1 & n_2^2 w^2 & n_1^2 w^3 & n_2^2 w^4
		\end{pmatrix}
		\,,
		\\
		\disc{\zeta \, w}
		&\doteq
		\begin{pmatrix}
			\frac{U^2}{a^2} w^1 &
			- \frac{W^2}{a^2} w^2 &
			\frac{U^2}{a^2} w^3 &
			- \frac{W^2}{a^2} w^4
		\end{pmatrix}
		\,.
	\end{aligned}
\end{equation}
Similar equations hold for arbitrary powers of $n$ and $\zeta$.

\subsubsection*{Symmetries of the discontinuity matrices}
Due to the electromagnetic symmetry, all discontinuity matrices are  of the form
\begin{equation}
	\label{eq:discontinuity matrix symmetries}
	\begin{pmatrix}
		a &
		b &
		n_1^2 e &
		n_2^2 f
		\\
		c &
		d &
		g &
		h
		\\
		-e &
		-f &
		a &
		b
		\\
		- g &
		- h &
		n_1^2 c &
		n_2^2 d
	\end{pmatrix}
	\,.
\end{equation}
Indeed, if the jump height of any component of the $\vec d$ field is given by
\begin{equation}
	\disc{d^i}
	\doteq
	\begin{pmatrix}
		\disc{d^i}^1 &
		\disc{d^i}^2 &
		\disc{d^i}^3 &
		\disc{d^i}^4
	\end{pmatrix}
	\,,
\end{equation}
we may apply the symmetry transformation $\vec d \mapsto \vec b$ and $\vec b \mapsto -\vec d/n^2$ to obtain
\begin{equation}
	\disc{b^i}
	\doteq
	\begin{pmatrix}
		-\disc{d^i}^3/n_1^2 &
		-\disc{d^i}^4/n_2^2 &
		\disc{d^i}^1 &
		\disc{d^i}^2
	\end{pmatrix}
	\,.
\end{equation}
Similarly, if the jump height of $e^i$ is given by
\begin{equation}
	\disc{e^i} \doteq
	\begin{pmatrix}
		\disc{e^i}^1 &
		\disc{e^i}^2 &
		\disc{e^i}^3 &
		\disc{e^i}^4
	\end{pmatrix}
	\,,
\end{equation}
we may apply the same transformation as above, which entails $\vec e \mapsto \vec h/n^2$, and find
\begin{equation}
	\disc{h^i} \doteq
	\begin{pmatrix}
		-\disc{e^i}^3 &
		-\disc{e^i}^4 &
		n_1^2 \disc{e^i}^1 &
		n_2^2 \disc{e^i}^2
	\end{pmatrix}
	\,.
\end{equation}
It thus suffices to compute the jump heights of $d^r$ and $e^z$, since the first two rows determine the entire matrix.

\subsubsection*{General Structure of Discontinuity matrices}

In general, there are two contributions to these matrices.
Some of the terms stem from the $z$ components of the fields, which produce terms in all other field components; while other terms stem from the cross product in \eqref{eq:rotation:matter fields} which relates $\vec d$ and $\vec b$ to $\vec e$ and $\vec h$.

Let $f$ be any function which enters the expressions for the $z$ components of the fields as
\begin{equation}
	\begin{aligned}
		d^z(r) = f(\alpha_1, \alpha_2; r) + \ldots \,,\\
		b^z(r) = f(\alpha_3, \alpha_4; r) + \ldots \,.
	\end{aligned}
\end{equation}
We denote by $\discM(f)$ the corresponding matrix defined by the condition that the such produced terms in $\disc{F}$ are
\begin{equation}
	\disc{F} = \discM(f) \bs \alpha + \ldots
\end{equation}
This method allows to describe all contributions to the discontinuity matrices which arise due to the $z$ components of the fields alone.
To cover the remaining terms which arise from the cross products in \eqref{eq:rotation:matter fields}, we will introduce additional matrices $\discM_1, \discM_2 $ etc.
Since, at our level of approximation, the cross product terms are determined by the unperturbed fields $\vec D$ and $\vec B$, which are parameterised by $\bs \alpha$ alone, these additional matrices will multiply $\bs \alpha$ only.

\subsection{Continuity of the Main Mode}

According to \eqref{eq:perturbation nomenclature}, \eqref{eq: unperturbed fundamental fields} and \eqref{eq:perturbed fundamental fields result}, the main Fourier modes of the $z$ components of $\vec d$ and $\vec b$ are given by
\begin{equation}
	\begin{aligned}
		d^z_0(r)
		&= D^z(r) + \delta d^z_0(r)
		= f_0(\alpha_1, \alpha_2; r)
		- 2 a^2 \beta \omega V^z p_0(\alpha_1, \alpha_2; r)
		\,,\\
		b^z_0(r)
		&= B^z(r) + \delta b^z_0(r)
		= f_0(\alpha_3, \alpha_4; r) 
		- 2 a^2 \beta \omega V^z p_0(\alpha_3, \alpha_4; r)
		\,.
	\end{aligned}
\end{equation}
Correspondingly, we write the jump height of the fields in the form
\begin{equation}
	\disc{F_0}
	=
	\discM(f_0) \bs \alpha
	- 2 a^2 \beta \omega V^z
	\left( \discM(p_0) + \discM_1 \right) \bs \alpha
	\,,
\end{equation}
where $\discM(f_0)$ describes the jump heights due to the functions $f_0$ alone, $\discM(p_0)$ those due to the presence of the function $p_0$, and $\discM_1$ the terms which arise form the cross product terms in \eqref{eq:rotation:matter fields}.
The factor $2 a^2 \beta \omega V^z$ was factored from $\discM_1$ for later convenience.

We start by showing that the matrix $\discM(f_0)$ is given by
\begin{equation}
	\label{eq:discontinuity matrix f_0}
	\begin{aligned}
		\discM(f_0) =
			&\begin{pmatrix}
				-i \frac{a \beta}{U} J_m' &
				-i \frac{a \beta}{W} K_m'	&
				m n_1^2 \frac{a \omega}{U^2} J_m &
				m n_2^2 \frac{a \omega}{W^2} K_m
			\\
				\frac{1}{n_1^2} J_m &
				- \frac{1}{n_2^2} K_m &
				0 &
				0
			\\
				- m \frac{a \omega}{U^2} J_m &
				- m \frac{a \omega}{W^2} K_m &
				-i \frac{a \beta}{U} J_m' &
				-i \frac{a \beta}{W} K_m'	&
			\\
				0 &
				0 &
				J_m &
				-K_m
			\end{pmatrix}
			\,,
	\end{aligned}
\end{equation}
where all functions whose arguments have been suppressed are evaluated at the core-cladding interface $r = a$. Thus, all \emph{ordinary} Bessel functions are evaluated at $U$ and all \emph{modified} Bessel functions are evaluated at $W$. Similarly, the functions $\Delta_\nu, T_\nu$ with suppressed arguments will be evaluated at $r/a =1$.

To compute this matrix, it suffices to consider the equations for the unperturbed fields, since the effects of all perturbations are described by $\discM(p_0)$ and $\discM_1$.
From \eqref{10IV19.1} we have
\begin{equation}
	i \zeta D^r = \beta \p_r D^z + i n^2 \omega \frac{m}{r} B^z
	\,.
\end{equation}
Using \eqref{eq:discontinuities unperturbed fields}, one obtains
\begin{equation}
	\begin{aligned}
		\disc{i \zeta D^r}
		&= \beta \disc{\p_r D^z} + i \frac{m \omega}{a} \disc{n^2 B^z}\\
		&\doteq \begin{pmatrix}
			\frac{\beta U}{a} J_m' &
			- \frac{\beta W}{a} K_m' &
			i n_1^2 \frac{m \omega}{a} J_m &
			-i n_2^2 \frac{m \omega}{a} K_m
		\end{pmatrix}
		\,,
	\end{aligned}
\end{equation}
and thus
\begin{equation}
	\disc{D^r}
	\doteq \begin{pmatrix}
		-i \frac{a \beta}{U} J_m' &
		-i \frac{a \beta}{W} K_m' &
		n_1^2 \frac{m a \omega}{U^2} J_m &
		n_2^2 \frac{m a \omega}{W^2} K_m
	\end{pmatrix}
	\,,
\end{equation}
which is exactly the first row of the matrix given in \eqref{eq:discontinuity matrix f_0}.
To compute the jump height of $E^z$, we use
\begin{equation}
	E^z = \frac{1}{n^2} D^z
	\,.
\end{equation}
By virtue of \eqref{eq:discontinuities unperturbed fields}, we find
\begin{equation}
	\disc{E^z} \doteq
	\begin{pmatrix}
		\frac{1}{n_1^2} J_m & - \frac{1}{n_2^2} K_m & 0 & 0 
	\end{pmatrix}
	\,,
\end{equation}
which is the second row of $\discM(f_0)$.
The third an fourth row are now determined by the electromagnetic symmetry, i.e.\ by the general form \eqref{eq:discontinuity matrix symmetries}.

To compute $\discM(p_0)$, we note that $\discM(f)$ is uniquely determined by the limiting behaviour of $f$ and its first derivative at $r = a$.
Comparing the limiting behaviour of $p_0$ and its first derivatives
with the limiting behaviour of $f_0$, we find that the matrix $\discM(p_0)$ is obtained from $\discM(f_0)$ by the substitution
\begin{equation}
	\begin{aligned}
		J_m(U)  &\mapsto \frac{\pi}{2} Y_m(U) \Delta_0(1)\,,\\
		J_m'(U) &\mapsto \frac{\pi}{2} Y_m'(U) \Delta_0(1)\,,\\
		K_m(U)  &\mapsto - I_m(U) T_0(1)\,,\\
		K_m'(U) &\mapsto - I_m'(U) T_0(1)\,.
	\end{aligned}
\end{equation}
We thus obtain
\begin{equation}
	\label{eq:discontinuity matrix p_0}
	\discM(p_0) =
	\begin{pmatrix}
		- i \frac{a \beta}{U} \frac{\pi}{2} Y_m' \Delta_0 &
		i \frac{a \beta}{W} I_m' T_0 &
		n_1^2 \frac{m a \omega}{U^2} \frac{\pi}{2} Y_m \Delta_0 &
		-  n_2^2 \frac{m a \omega}{W^2} I_m T_0
	\\
		\frac{1}{n_1^2} \frac{\pi}{2} Y_m \Delta_0 &
		\frac{1}{n_2^2} I_m T_0 &
		0 &
		0
	\\
		- \frac{m a \omega}{U^2} \frac{\pi}{2} Y_m \Delta_0 &
		\frac{m a \omega}{W^2} I_m T_0 &
		- i \frac{a \beta}{U} \frac{\pi}{2} Y_m' \Delta_0 &
		i \frac{a \beta}{W} I_m' T_0
	\\
		0 &
		0 &
		\frac{\pi}{2} Y_m \Delta_0 &
		I_m T_0
	\end{pmatrix}
	\,.
\end{equation}

Finally, we compute $\discM_1$ explicitly.
According to \eqref{eq: perturbed components A (m mode)}, the relevant term for $\delta d^z_0$ is
\begin{equation}
	i \zeta \delta d^r_0
	= - i V^z \beta (\beta H^\theta + \omega D^r)
	+ \ldots
\end{equation}
From \eqref{10IV19.1} we obtain
\begin{equation}
	\label{eq:discontinuity intermediate 1}
	i \zeta (\beta H^\theta + \omega D^r)
	= 2 \beta \omega \p_r D^z + i \bar \zeta \frac{m}{r} B^z
	\,,
\end{equation}
where we have set
\begin{equation}
	\bar \zeta
	= n^2 \omega^2 + \beta^2
	\equiv \zeta + 2 \beta^2
	=\begin{cases}
		\bar U^2/a^2, & r < a\,,\\
		\bar W^2/a^2, & r > a\,.
	\end{cases}
\end{equation}
Here, we have introduced the constants $\bar U$ and $\bar W$ according to
\begin{equation}
	\begin{aligned}
		\bar U^2/a^2 &= \beta^2 + n_1^2 \omega^2 \equiv  2 \beta^2 + U^2/a^2\,,\\
		\bar W^2/a^2 &= \beta^2 + n_2^2 \omega^2 \equiv  2 \beta^2 - W^2/a^2\,.
	\end{aligned}
\end{equation}
Having established that
\begin{equation}
	\zeta^2 \delta d^r_0
	= i V^z \beta \left(
	2 \beta \omega \p_r D^z + i \bar \zeta \frac{m}{r} B^z	
	\right)
	+ \ldots
	\,,
\end{equation}
we may use \eqref{eq:discontinuities unperturbed fields} to obtain
\begin{equation}
	\disc{\zeta^2 \delta d^r_0}
	\doteq i V^z \beta
	\begin{pmatrix}
		2 \beta \omega \frac{U}{a} J_m' &
		- 2 \beta \omega \frac{W}{a} K_m' &
		i \frac{m \bar U^2}{a^3} J_m &
		- i \frac{m \bar W^2}{a^3} K_m
	\end{pmatrix}_{\bs \alpha}
	+ \ldots
	\,,
\end{equation}
from which it follows that
\begin{equation}
	\disc{\delta d^r_0}
	\doteq
	- 2 a^2 \beta \omega V^z
	\begin{pmatrix}
		- i \frac{a \beta}{U^3} J_m' &
		+ i \frac{a \beta}{W^3} K_m' &
		+ \frac{m}{2 a \omega} \frac{\bar U^2}{U^4} J_m &
		- \frac{m}{2 a \omega} \frac{\bar W^2}{W^4} K_m
	\end{pmatrix}_{\bs \alpha}
	+ \ldots
	\,.
\end{equation}
This determines the first row of $\discM_1$.
According to \eqref{eq: perturbed components B (m mode)} we have $n^2 \delta e^z_0 = \delta d^z_0$, so $\vec V$ does not contribute to $\delta e^z_0$. Accordingly, the second row of $\discM_1$ vanishes.
The remaining rows are then then fully determined by the general form  \eqref{eq:discontinuity matrix symmetries},
so we arrive at the result
\begin{equation}
	\label{eq:discontinuity matrix D_1}
	\discM_1
	= \begin{pmatrix}
		- i \frac{a \beta}{U^3} J_m' &
		i \frac{a \beta}{W^3} K_m' &
		\frac{m}{2 a \omega} \frac{\bar U^2}{U^4} J_m &
		-  \frac{m}{2 a \omega} \frac{\bar W^2}{W^4} K_m
	\\
		0 &
		0 &
		0 &
		0
	\\
		- \frac{m}{2 a \omega} \frac{1}{n_1^2} \frac{\bar U^2}{U^4} J_m &
		\frac{m}{2 a \omega} \frac{1}{n_2^2} \frac{\bar W^2}{W^4} K_m &
		- i \frac{a \beta}{U^3} J_m' &
		i \frac{a \beta}{W^3} K_m'
	\\
		0 &
		0 &
		0 &
		0
	\end{pmatrix}
	\,.
\end{equation}
The continuity conditions of the main mode now take the concise form
\begin{highlight}
\begin{equation}
	\label{eq:continuity main mode}
	\disc{F_0}
	= \discM(f_0) \bs \alpha
	- 2 a^2 \beta \omega V^z \left( \discM(p_0) + \discM_1 \right) \bs \alpha
	\overset ! = 0\,.
\end{equation}
\end{highlight}

\subsection{Dispersion Relation}

We have established that the continuity condition of the main mode takes the form
\begin{equation}
	\label{eq:continuity main mode abstract}
	(M^{(0)} + \delta M) \bs \alpha = 0\,,
\end{equation}
where
\begin{equation}
	\begin{aligned}
		M^{(0)}
		&= \discM(f_0)\,,
		\\
		\delta M
		&= -2 a^2 \beta \omega V^z \left( \discM(p_0) + \discM_1 \right)
		\,.
	\end{aligned}
\end{equation}
In the unperturbed case, where $\Omega = 0$, $M$ reduces to $M^{(0)}$, so $\delta M$ is the first order perturbation of $M$ due to Earth's rotation.

Generically, the kernel of the matrix $M$ is trivial, so we must arrange parameters such that its determinant vanishes, if we wish to obtain a nontrivial solution.
Since the only remaining parameters are $\beta$ and $\omega$, one is determined by the other via the dispersion relation
\begin{highlight}
\begin{equation}
	\label{eq:dispersion:exact}
	\det M = \det(M^{(0)} + \delta M) \overset ! = 0\,.
\end{equation}
\end{highlight}
We choose to take the frequency $\omega$ as the independent variable, such that the dispersion relation determines $\beta = \beta(\omega)$.
Equation \eqref{eq:dispersion:exact} is of major importance since encodes how the wave vector $\beta$ changes due to Earth’s rotation.
As in the unperturbed case, this equation can only be solved numerically.
If a solution to the unperturbed dispersion relation is known, one may obtain the first order correction to $\beta$ by expanding the dispersion relation \eqref{eq:dispersion:exact} to first order in the velocity $V$.

Here, we see the importance of letting the parameters $\bs \alpha$ deviate from their unperturbed values $\bs \alpha^{(0)}$. Had we added homogeneous solutions to the wave equations with parameters $\bs \alpha^{(1)}$ and neglected the term $\delta M \bs \alpha^{(1)}$, which is quadratic in the velocity $\vec V$, instead of \eqref{eq:continuity main mode abstract} we would have obtained
\begin{equation}
	M^{(0)} \bs \alpha^{(1)} + \delta M \bs \alpha^{(0)} = 0\,,
\end{equation}
since $M^{(0)} \bs \alpha^{(0)} = 0$.
This equation already suggests that the dispersion relation must be modified since it cannot be solved for $\bs \alpha^{(1)}$ due to $M$ being singular when $\beta$ assumes its unperturbed value.
Further analysis would have been necessary to obtain the dispersion relation \eqref{eq:dispersion:exact}.

Linearising the condition $\det M = 0$ in the vicinity of $V = 0$ using Jacobi's formula
\begin{equation}
	\frac{\p}{\p A_{ij}} \det A = \adj(A)^{ji} \,,
\end{equation}
where $\adj(A)$ is the adjugate matrix of $A$ (the transpose of the cofactor matrix), one obtains
\begin{equation}
	\delta (\det M)
	=
	\underbrace{
	 \left[
		\frac{\p |M^{(0)}|}{\p \beta}
		+ \frac{\p |M^{(0)}|}{\p U} \frac{\p U}{\p \beta}
		+ \frac{\p |M^{(0)}|}{\p W} \frac{\p W}{\p \beta}
	\right]	}_{\equiv \frac{\dd}{\dd \beta} \det M^{(0)} } \delta \beta
	+ \tr \left[ \adj(M^{(0)}) \, \delta M \right] = 0\,,
\end{equation}
which implies
\begin{equation}
	\delta \beta
	= - \frac{\tr \left[ \adj(M^{(0)}) \, \delta M \right]}{\frac{\dd}{\dd \beta} \det M^{(0)}}
	\,.
\end{equation}
When solving the equation $\det M = 0$, we may equally replace $M$ by $\tilde M = A_1 M A_2$, where
\begin{equation}
	A_1 = \begin{pmatrix}
		0 &
		1 &
		0 &
		0
	\\
		1 &
		i n_1^2 \frac{a \beta}{U} \frac{J'_m}{J_m} &
		0 &
		- n_1^2 \frac{m a \omega}{U^2}
	\\
		0 &
		0 &
		0 &
		1
	\\
		0 &
		n_1^2 \frac{m a \omega}{U^2} &
		1 &
		i \frac{a \beta}{U} \frac{J_m'}{J_m}
	\end{pmatrix},
	\qquad
	A_2 =
	\begin{pmatrix}
		n_1^2 &
		0 &
		0 &
		0
	\\
		0 &
		n_2^2 &
		0 &
		0
	\\
		0 &
		0 &
		1 &
		0
	\\
		0 &
		0 &
		0 &
		1
	\end{pmatrix}.
\end{equation}
Since these matrices are nonsingular (in fact $\det A_1 = 1$ and $\det A_2 = n_1^2 n_2^2$) this does not change the roots of $\det M = 0$.

It proves useful to introduce the abbreviations
\begin{equation}
	\begin{aligned}
		\Delta n^2
		&= n_1^2 - n_2^2 \,,
		\\
		\nu &=
		\frac{a^2 \beta^2}{U^2 W^2} m \Delta n^2 \omega \,,
	\end{aligned}
\end{equation}
as well as the brackets
\begin{equation}
	\begin{aligned}
		\braketP{J_m}{K_m} &= \frac{1}{U} J'_m(U) K_m(W) + \frac{1}{W} J_m(U) K_m'(W) \,,\\
		\braketM{J_m}{K_m} &= \frac{1}{U^3} J_m'(U) K_m(W) - \frac{1}{W^3} J_m(U) K_m'(W) \,,\\
		\braketP{J_m}{K_m}_{n^2} &= \frac{n_1^2}{U} J'_m(U) K_m(W) + \frac{n_2^2}{W} J_m(U) K_m'(W) \,,\\
		\braketM{J_m}{K_m}_{n^2} &= \frac{n_1^2}{U^3} J_m'(U) K_m(W) - \frac{n_2^2}{W^3} J_m(U) K_m'(W) \,.
	\end{aligned}
\end{equation}
$\tilde M^{(0)}$ then takes the concise form
\ptcr{cool}
\begin{equation}
	\tilde M^{(0)}
	=
	\begin{pmatrix}
		J_m &
		- K_m &
		0 &
		0
	\\
		0 &
		-i \frac{a \beta}{J_m} \braketP{J_m}{K_m}_{n^2} &
		0 &
		a \nu K_m
	\\
		0 &
		0 &
		J_m &
		-K_m
	\\
		0 &
		- a \nu K_m &
		0 &
		-i \frac{a \beta}{J_m} \braketP{J_m}{K_m}
	\end{pmatrix} \,,
\end{equation}
from which one can read off the determinant
\begin{equation}
	\det \tilde M^{(0)} =
	a^2 \nu^2 J_m(U)^2 K_m(W)^2
	- a^2 \beta^2 \braketP{J_m}{K_m} \braketP{J_m}{K_m}_{n^2} \,.
\end{equation}
Since similar structures will appear again in the following calculations, we set
\begin{equation}
	\Psi(f, g) := a^2 \nu^2 J_m(U)^2 f(W) g(W) - a^2 \beta^2 \braketP{J_m}{f} \braketP{J_m}{g}_{n^2} \,,
\end{equation}
such that the defining equation of the unperturbed value of $\beta$ takes the form
\begin{equation}
	\label{eq:dispersion:unperturbed}
	\det \tilde M^{(0)} = \Psi(K_m, K_m) = 0 \,.
\end{equation}
Using the fact that
\begin{equation}
	\frac{n_1^2}{U^2} + \frac{n_2^2}{W^2} = \frac{a^2 \beta^2}{U^2 W^2} \Delta n^2 \,,
\end{equation}
as well as the Wronskian
\begin{equation}
	J_m(U) Y_m'(U) - J_m'(U) Y_m(U) = \frac{2}{\pi U} \,,
\end{equation}
the adjugate of $\tilde M^{(0)}$ and the perturbations $\delta \tilde M_1$, $\delta \tilde M_2$ take the form
\begin{equation}
	\begin{aligned}
		\adj ( \tilde M^{(0)})
			=
			&\begin{pmatrix}
				0 &
				-i a \beta \braketP{J_m}{K_m} K_m
				&
				0
				&
				- a \nu J_m K_m^2
			\\
				0 &
				-i a \beta \braketP{J_m}{K_m} J_m &
				0 &
				- a \nu J_m^2 K_m
			\\
				0 &
				a \nu J_m K_m^2  &
				0 &
				-i a \beta \braketP{J_m}{K_m}_{n^2} K_m
			\\
				0 &
				a \nu J_m^2 K_m &
				0 &
				-i a \beta \braketP{J_m}{K_m}_{n^2} J_m
			\end{pmatrix}
		\,,
		\\
		\delta \tilde M_1
			= - 2 a^2 \beta \omega V^z
			&\begin{pmatrix}
				\frac{\pi}{2} Y_m \Delta_0 &
				I_m T_0 &
				0 &
				0
			\\
				- i \frac{n_1^2}{U^2} \frac{a \beta}{J_m} \Delta_0 &
				i \frac{a \beta}{J_m} \braketP{J_m}{I_m}_{n^2} T_0 &
				0 &
				- a \nu I_m T_0
			\\
				0 &
				0 &
				\frac{\pi}{2} Y_m \Delta_0 &
				I_m T_0
			\\
				0 &
				a \nu I_m T_0 &
				-i \frac{1}{U^2} \frac{a \beta}{J_m} \Delta_0 &
				i \frac{a \beta}{J_m} \braketP{J_m}{I_m} T_0
			\end{pmatrix}
		\,,
		\\
		\delta \tilde M_2
			= -2 a^2 \beta \omega V^z
			&\begin{pmatrix}
				0 &
				0 &
				0 &
				0
			\\
				-i n_1^2 \frac{a \beta}{U^3} J_m' &
				i n_2^2 \frac{a \beta}{W^3} K_m' &
				\frac{m}{2 a \omega} \frac{\bar U^2}{U^4} J_m &
				-\frac{m}{2 a \omega} \frac{\bar W^2}{W^4} K_m
			\\
				0 &
				0 &
				0 &
				0
			\\
				- \frac{m}{2 a \omega} \frac{\bar U^2}{U^4} J_m &
				\frac{m}{2 a \omega} \frac{\bar W^2}{W^4} K_m &
				- i \frac{a \beta}{U^3} J_m' &
				i \frac{a \beta}{W^3} K_m'
			\end{pmatrix}
		\,.
	\end{aligned}
\end{equation}
From these matrices, one obtains the following expressions for their traces with $\tilde M^{(0)}$:
\begin{equation}
	\begin{aligned}
		\frac{\tr[\tilde M^{(0)} \delta \tilde M_1]}{2 a^2 \beta \omega V^z}
		&=
		\frac{a^2 \beta^2}{U^2} \frac{K}{J} \left( n_1^2 \braketP{J}{K} + \braketP{J}{K}_{n^2} \right) \Delta_0
		+ \left( \Psi(K, I) + \Psi(I, K) \right) T_0 \,,
		\\
		\frac{\tr[\tilde M^{(0)} \delta \tilde M_2]}{2 a^2 \beta \omega V^z}
		&=
		a^2 \beta^2 \left( \braketP{J}{K} \braketM{J}{K}_{n^2} + \braketP{J}{K}_{n^2} \braketM{J}{K} \right)
		+ \frac{m \nu}{a^2 \omega} J^2 K^2 \left( \frac{\bar W^2}{W^4} - \frac{\bar U^2}{U^4} \right) \,.
	\end{aligned}
\end{equation}
We thus arrive at the following result for the linear change in $\beta$:
%
%
\begin{highlight}
\begin{equation}
	\label{eq:dispersion:linearised result}
	\delta \beta = - \frac{2 a^2 \beta \omega V^z}{\frac{\dd}{\dd \beta} \det \tilde M^{(0)}}
		\bigg[ A + B + C + D \bigg]\,,
\end{equation}
\end{highlight}
where
\begin{equation}
	\begin{aligned}
		A &=
		\left( \Psi(K_m, I_m) + \Psi(I_m, K_m) \right) T_0(1)
		\,,\\
		B &=
			\frac{a^2 \beta^2}{U^2} \frac{K_m(W)}{J_m(U)}
			\left( n_1^2 \braketP{J_m}{K_m} + \braketP{J_m}{K_m}_{n^2} \right) \Delta_0(1)
		\,,\\
		C &=
			a^2 \beta^2 \left( \braketP{J_m}{K_m} \braketM{J_m}{K_m}_{n^2} + \braketP{J_m}{K_m}_{n^2} \braketM{J_m}{K_m} \right)
		\,,\\
		D &=
		a^2 \beta^2 m^2 \Delta n^2 \frac{J_m(U)^2 K_m(W)^2}{U^2 W^2} \left( \frac{\bar W^2}{W^4} - \frac{\bar U^2}{U^4} \right)
		\,.
	\end{aligned}
\end{equation}
This formula is the main result of this section: it determines the linearised deviation from the unperturbed dispersion relation \eqref{eq:dispersion:unperturbed}. If $\beta$ solves the unperturbed equation for given $\omega$, \eqref{eq:dispersion:linearised result} determines the first order correction to $\beta$ for constant $\omega$.

Once the perturbed value of $\beta$ is known, the coefficients $\bs \alpha$ (which parameterise the main mode) are obtained by finding the kernel of the matrix
\begin{equation}
	M = M^{(0)} + \delta M \,,
\end{equation}
cf.~\eqref{eq:continuity main mode} and \eqref{eq:continuity main mode abstract}.
From the unperturbed problem it is known that the kernel of $M^{(0)}$ is one-dimensional.
Since the dimension of the kernel is an upper semi-continuous function, the same will hold for $M$, provided that the velocity $V$ is sufficiently small.
Note that the determination of $\bs \alpha$ to order $\order(V)$ requires $M^{(0)}$ to be computed using the perturbed value of $\beta$. Since $\delta M$ is already of order $\order(V)$ one may use the unperturbed value of $\beta$ in this matrix.

\subsection{Continuity of the Sidebands}

In this section, we compute the discontinuity matrices associated to the sidebands.

From \eqref{eq:perturbed fundamental fields result} it is clear that the jump height of the sidebands takes the form
\begin{equation}
	\label{eq:continuity sidebands}
	\disc{F_\pm}
	=
	\discM(f_\pm) \bs \chi^\pm
	- 2 a \omega V^\pm \discM(p_\pm) \bs \alpha
	- i V^\pm \discM_2^\pm \bs \alpha
	\,.
\end{equation}
The coefficient in front of $\discM_2^\pm$, which contains the contributions from the usual cross product terms, was factored for convenience.

We may obtain $\discM(f_\pm)$ from $\discM(f_0)$ by noting that
\begin{equation}
	f_\pm(\alpha_1, \alpha_2; r)
	= \begin{cases}
		\mp U J_{m \pm 1}(U r/a) & r < a\,,\\
		- W K_{m \pm 1}(W r/a) & r > a\,,\\
	\end{cases}
\end{equation}
is related to $f_0(\alpha_1, \alpha_2; r)$ by the substitution
\begin{equation}
	\begin{aligned}
		J_m & \mapsto \pm U J_{m \pm 1}
		\,,\\
		K_m & \mapsto - W K_{m \pm 1}
		\,.
	\end{aligned}
\end{equation}
Applying this to $\discM(f_0)$ and replacing all factors $m$ (which are due to derivatives with respect to the angle $\theta$) by $m \pm 1$, we find
\begin{equation}
	\discM(f_\pm)
	=
	\begin{pmatrix}
		\pm i a \beta J'_{m \pm 1} &
		i a \beta K'_{m \pm 1} &
		\mp n_1^2 \frac{(m \pm 1) a \omega}{U} J_{m \pm 1} &
		- n_2^2 \frac{(m \pm 1) a \omega}{W} K_{m \pm 1}
		\\
		\mp \frac{U}{n_1^2} J_{m \pm 1} &
		\frac{W}{n_2^2} K_{m \pm 1} &
		0 &
		0
		\\
		\pm \frac{(m \pm 1) a \omega}{U} J_{m \pm 1} &
		\frac{(m \pm 1) a \omega}{W} K_{m \pm 1} &
		\pm i a \beta J'_{m \pm 1} &
		i a \beta K'_{m \pm 1}
		\\
		0 &
		0 &
		\mp U J_{m \pm 1} &
		W K_{m \pm 1}
	\end{pmatrix}
	\,.
\end{equation}
Similarly, comparing $p_0$ with $p_\pm$ in \eqref{eq: particular solutions}, find that $\discM(p_\pm)$ is related to $\discM(p_0)$ by the substitution
\begin{equation}
	\begin{aligned}
		m &\mapsto m \pm 1\,,\\
		\Delta_0 &\mapsto \mp \Delta_\pm \,,\\
		T_0 &\mapsto - T_\pm \,,
	\end{aligned}
\end{equation}
so we find
%
\begin{equation}
	\discM(p_\pm)
	=
	\begin{pmatrix}
		\pm i a \beta Y'_{m \pm 1} &
		-i a \beta I'_{m \pm 1} &
		\mp n_1^2 \frac{(m \pm 1) a \omega}{U} Y_{m \pm 1} &
		n_2^2 \frac{(m \pm 1) a \omega}{W} I_{m \pm 1} 
		\\
		\mp \frac{U}{n_1^2} Y_{m \pm 1} &
		-\frac{W}{n_2^2} I_{m \pm 1} &
		0 &
		0
		\\
		\pm \frac{(m \pm 1) a \omega}{U} Y_{m \pm 1} &
		-\frac{(m \pm 1) a \omega}{W} I_{m \pm 1} &
		\pm i a \beta Y'_{m \pm 1} &
		-i a \beta I'_{m \pm 1} 
		\\
		0 &
		0 &
		\mp U Y_{m \pm 1} &
		- W I_{m \pm 1}
	\end{pmatrix}
	\Lambda^\pm(\Delta, T)
	\,,
\end{equation}
where we have set
\begin{equation}
	\Lambda^\pm(\Delta, T)
	= \diag\left(
		\frac{\pi}{2} \Delta_\pm(1),
		T_\pm(1),
		\frac{\pi}{2} \Delta_\pm(1),
		T_\pm(1)
	\right)
	\,.
\end{equation}

Finally, we compute the matrix $\discM_2^\pm$.
To this end, we note that the relevant terms in the equation for $\delta d^r_\pm$ are
\begin{equation}
	i \zeta \delta d^r_\pm
	= - i V^\pm c_m^\pm( \beta H^\theta + \omega D^r)
	\pm \frac{V^\pm}{r} (\beta H^r - \omega D^\theta)
	+ \ldots
\end{equation}
Using \eqref{eq:discontinuity intermediate 1} and the analogous equation
\begin{equation}	
	\label{eq:discontinuity intermediate 2}
	i \zeta (\beta H^r - \omega D^\theta)
	= \bar \zeta \p_r B^z - 2 i \beta \omega \frac{m}{r} D^z
	\,,
\end{equation}
we obtain
\begin{equation}
	\zeta^2 \delta d^r_\pm
	=
	  2 i \beta \omega V^\pm \left(
	  	c_m^\pm(\p_r D^z)
	  	\pm \frac{m}{r^2} D^z
	  \right)
	- \bar \zeta V^\pm \left(
		c_m^\pm \left( \frac{m}{r} B^z \right)
		\pm \frac{1}{r} \p_r B^z
	\right)
	\,.
\end{equation}
Simplifying this expression using the identities
\begin{equation}
	\begin{aligned}
		\p_r( c_m^\pm f)
		&= c_m^\pm (\p_r f) \pm \frac{m}{r^2} f
		\,,\\
		\frac{m \pm 1}{r}	c_m^\pm(f)
		&= c_m^\pm\left( \frac{m}{r} f \right) \pm \frac{1}{r} \p_r f
		\,,
	\end{aligned}
\end{equation}
one obtains
\begin{equation}
	\zeta^2 \delta d^r_\pm
	=  2 i \beta \omega V^\pm \p_r (c_m^\pm D^z)
	- \bar \zeta V^\pm \frac{m \pm 1}{r} c_m^\pm B^z
	\,.
\end{equation}
Using the explicit expressions for $D^z$ and $B^z$, as well as the recursion relations for the Bessel functions \eqref{eq:Bessel recursion}, we find
\begin{equation}
	\begin{aligned}
		\disc{\p_r c_m^\pm D^z}
		&\doteq
		\begin{pmatrix}
			\mp \frac{U^2}{a^2} J'_{m \pm 1} &
			\frac{W^2}{a^2} K'_{m \pm 1} &
			0 &
			0
		\end{pmatrix}
		\,,
		\\
		\disc{\bar \zeta c_m^\pm B^z}
		&\doteq
		\begin{pmatrix}
			0 &
			0 &
			\mp \frac{U \bar U^2}{a^3} J_{m \pm 1} &
			\frac{W \bar W^2}{a^3} K_{m \pm 1}
		\end{pmatrix}
		\,,
	\end{aligned}
\end{equation}
which implies
\begin{equation}
	\disc{\delta d^r_\pm}
	\doteq
	- i V^\pm
	\begin{pmatrix}
		\pm \frac{2 a^2 \beta \omega}{U^2} J'_{m \pm 1} &
		- \frac{2 a^2 \beta \omega}{W^2} K'_{m \pm 1} &
		\pm i \frac{m \pm 1}{U} \frac{\bar U^2}{U^2} J_{m \pm 1} &
		- i \frac{m \pm 1}{W} \frac{\bar W^2}{W^2} K_{m \pm 1}
	\end{pmatrix}
	\,.
\end{equation}
This determines the first row of $\discM_2^\pm$.

Next, we compute the second row, which is (up to the factor $- i V^\pm$) given by the jump height of $\delta e^z_\pm$.
From \eqref{eq: perturbed components B (sidebands)} we find the relevant term to be
\begin{equation}
	\delta e^z_\pm = -i V^\pm (B^\theta \mp i B^r)
	\,.
\end{equation}
From \eqref{10IV19.1} we have
\begin{equation}
	i \zeta (B^\theta \mp i B^r)
	= c_m^\pm ( \omega D^z \mp i \beta B^z)
	\,.
\end{equation}
Substituting this into the above expression for $\delta e^z_\pm$ and computing the jump height, we find
\begin{equation}
	\disc{n^2 \zeta \delta e^z_\pm}
	\doteq -V^\pm
	\begin{pmatrix}
		\mp \frac{\omega U}{a} J_{m \pm 1} &
		\frac{\omega W}{a} K_{m \pm 1} &
		i \frac{\beta U}{a} J_{m \pm 1} &
		\mp i \frac{\beta W}{a} K_{m \pm 1}
	\end{pmatrix}
	\,,
\end{equation}
which implies
\begin{equation}
	\disc{\delta e^z_\pm}
	\doteq -i V^\pm
	\begin{pmatrix}
		\pm \frac{i}{n_1^2} \frac{a \omega}{U} J_{m \pm 1} &
		\frac{i}{n_2^2} \frac{a \omega}{W} K_{m \pm 1} &
		\frac{1}{n_1^2} \frac{a \beta}{U} J_{m \pm 1} &
		\pm \frac{1}{n_2^2} \frac{a \beta}{W} K_{m \pm 1}
	\end{pmatrix}
	\,.
\end{equation}
Using the general form \eqref{eq:discontinuity matrix symmetries}, we thus arrive at the result
\begin{equation}
	\discM_2^\pm
	=
	\begin{pmatrix}
		\pm \frac{2 a^2 \beta \omega}{U^2} J'_{m \pm 1} &
		- \frac{2 a^2 \beta \omega}{W^2} K'_{m \pm 1} &
		\pm i \frac{m \pm 1}{U} \frac{\bar U^2}{U^2} J_{m \pm 1} &
		- i \frac{m \pm 1}{W} \frac{\bar W^2}{W^2} K_{m \pm 1}
		\\
		\pm \frac{i}{n_1^2} \frac{a \omega}{U} J_{m \pm 1} &
		\frac{i}{n_2^2} \frac{a \omega}{W} K_{m \pm 1} &
		\frac{1}{n_1^2} \frac{a \beta}{U} J_{m \pm 1} &
		\pm \frac{1}{n_2^2} \frac{a \beta}{W} K_{m \pm 1}
		\\
		\mp \frac{i}{n_1^2} \frac{m \pm 1}{U} \frac{\bar U^2}{U^2} J_{m \pm 1} &
		\frac{i}{n_2^2} \frac{m \pm 1}{W} \frac{\bar W^2}{W^2} K_{m \pm 1} &
		\pm \frac{2 a^2 \beta \omega}{U^2} J'_{m \pm 1} &
		- \frac{2 a^2 \beta \omega}{W^2} K'_{m \pm 1}
		\\
		- \frac{1}{n_1^2} \frac{a \beta}{U} J_{m \pm 1} &
		\mp \frac{1}{n_2^2} \frac{a \beta}{W} K_{m \pm 1} &
		\pm i \frac{a \omega}{U} J_{m \pm 1} &
		i \frac{a \omega}{W} K_{m \pm 1}
	\end{pmatrix}
	\,.
\end{equation}

Having computed all matrices which enter \eqref{eq:continuity sidebands} and having computed the coefficients $\bs \alpha$ of the main mode, the parameters $\bs \chi^\pm$ are determined by
\begin{highlight}
\begin{equation}
	\label{eq:continuity condition sidebands}
	\disc{F_\pm}
	=
	\discM(f_\pm) \bs \chi^\pm
	- 2 a \omega V^\pm \discM(p_\pm) \bs \alpha
	- i V^\pm \discM_2^\pm \bs \alpha
	\overset ! = 0
	\,.
\end{equation}
\end{highlight}
Here and in the following sections, we shall assume that the matrices $\discM(f_\pm)$ are invertible.
Recall that the unperturbed dispersion relation is such that $\det \discM(f_0) = 0$.
The special case where both $\discM(f_0)$ and either one of the two matrices $\discM(f_\pm)$ are singular for the same value of $\beta$ would require separate analysis.

\section{Further Corrections}
\label{s:further corrections}

In this section, we briefly discuss additional corrections of order $\ell \Omega$ and $\lambda \Omega$, where $\ell$ is the length of the waveguide and $\lambda$ the light wavelength.
Corrections of order $\lambda \Omega$ arise from the first order differential operator in equation \eqref{eq:wave:full}, and corrections of order $\ell \Omega$ are obtained by improving the approximation $\vec x = \vec R + \vec r \approx \vec R$ to $\vec x \approx \vec R + z \vec e_z$.
By adding these two corrections, the only neglected contribution of order $\Omega$ is of order $\order(\Omega a)$, where $a$ is the core radius of the dielectric.

Due to their small amplitudes, we expect that the sidebands will not be experimentally accessible in the near future and thus restrict our attention to the main mode.
We show that this main mode is unaffected by the above mentioned effects, i.e.\ the contributions of order $\ell \Omega$ and $\lambda \Omega$ vanish.

\subsection{Corrections of Order \texorpdfstring{$\lambda \Omega$}{λΩ}}
\label{s:correction:wavelength}

In the previous analysis, the source term proportional to $\Omega$ in \eqref{eq:wave:full} was neglected.
In this section, we briefly discuss the influence of this term and thus consider the equations
\begin{equation}
	\begin{aligned}
		\mu ( n^2 \delta \ddot d^z - \Delta \delta d^z)
		&= + 2 \Omega(B^z) + 2 i \beta \vec \Omega \cdot \vec B
		\,,\\
		\varepsilon( n^2 \delta \ddot b^z - \Delta \delta b^z)
		&= - 2 \Omega(D^z) - 2 i \beta \vec \Omega \cdot \vec D
		\,.
	\end{aligned}
\end{equation}
The additional terms do not modify the main mode, since the $m$-th Fourier component vanishes:
\begin{equation}
	\braket{\Omega(D^z) + i \beta \vec \Omega \cdot \vec D}_0
	= \Omega^z \p_z D^z + i \beta \Omega^z D^z
	= 0
	\,.
\end{equation}
Thus, this correction does not modify our result for the shift in $\beta$ but merely modifies the coefficients $\chi^\pm_1, \ldots, \chi^\pm_4$, which parameterise the sidebands according to \eqref{eq:perturbed fundamental fields result}.

\subsection{Corrections of Order \texorpdfstring{$\ell \Omega$}{lΩ}}
\label{s:correction:length}

Hitherto, we have approximated
\begin{equation}
	\vec x = \vec R + \vec r \approx \vec R \,,
\end{equation}
to obtain the simplified wave equation \eqref{eq:wave:approximate} and to compute the transverse field components, e.g.~in \eqref{eq: perturbed components A (m mode)} and \eqref{eq: perturbed components B (m mode)}.
This approximation is expected to produce relative errors of the order of $a/R$ or $\ell/R$, where $a$ is the radius of the dielectric, $\ell$ its length and $R$ is Earth’s radius.
Instead of the above approximation, we thus write
\begin{equation}
	\vec x
	= \vec R + \vec r
	\approx R + z \vec e_z
	\,.
\end{equation}
The corresponding approximation for the velocity vector field $\vec v$ reads
\begin{equation}
	\label{eq:l/R:v approx}
	\vec v
	= \vec V + \vec \Omega \times \vec r
	\approx \vec V + \vec u
	\,,
\end{equation}
where the relative velocity $\vec u = \vec \Omega \times (z \vec e_r)$ has the following components when referred to a cylindrical orthonormal frame:
\begin{equation}
	u^r = + \Omega^\theta z\,,
	\qquad
	u^\theta = - \Omega^r z\,,
	\qquad
	u^z = 0\,.
\end{equation}
But since $\vec u$ depends only on $\Omega^r$ and $\Omega^\theta$, but not on $\Omega^z$, it depends on the angle $\theta$ by $\sin \theta$ or $\cos \theta$, but it does not have a constant $\theta$-independent component.
Hence, all terms arising from $\vec u$ only affect the sidebands and leave the main mode unchanged.
We thus conclude that the entire analysis of the continuity conditions for the main mode is unaltered by the terms introduced here.

\section{Numerical Examples}
\label{chapter:application}

We apply our formulae to the same setup as considered in \cite{Beig_2018}. Table \ref{table:parameters} lists both the parameters of the waveguide (refractive indices and core diameter) and the parameters of the light wave (wavelength and angular mode number).
\begin{table}[h]
	\centering
	\begin{tabular}{@{}lll@{}}
		\toprule
			& Quantity
			& Magnitude
		\\
		\midrule
			$n_1$
			& Core refractive index
			& \SI{1.4712}{}
		\\
			$n_2$
			& Cladding refractive index
			& \SI{1.4659}{}
		\\
			$n$
			& Effective refractive index
			& \SI{1.4682}{}
		\\
			$a$
			& Core diameter
			& \SI{4.1}{\micro\meter}
		\\
			$\lambda_0$
			& Wavelength in vacuum
			& \SI{1550}{\nano\meter}
		\\
			$m$
			& Angular mode number
			& \SI{1}{}
		\\
		\bottomrule
	\end{tabular}
	\caption{Parameters used for numerical examples.}
	\label{table:parameters}
\end{table}

\noindent
The effective refractive index $n$ was found by solving \eqref{eq:dispersion:unperturbed} numerically, which yields
\begin{equation}
	\beta \approx 1.4682 \, \omega
	\,.
\end{equation}
%

\subsection{Corrections to the Dispersion Relation}

The results derived in this work may be applied both to Earth’s rotation about its own axis (spin) and its orbital rotation about the Sun, if the eccentricity of Earth’s orbit is neglected.

In the first case, we use $R = R_E \approx \SI{6371}{\kilo\meter}$ and $\Omega = \SI[per-mode=repeated-symbol]{2 \pi}{\per\day}$.
Since Earth is slightly oblate, more accurate results could be obtained by setting $R$ to be the local distance to Earth’s axis.
To illustrate the dependence of the effect on the latitude $\psi$, we compute $\delta \beta$ both at the equator ($\psi = \pi/2$) and in Vienna ($\psi = \ang{48;12;30}$). In both cases, we assume the waveguide to be aligned with the velocity vector due to the rotation, i.e.\ the $z$ axis points east.

To estimate the effect of Earth’s orbital motion, we set $\Omega = \SI[per-mode=repeated-symbol]{2\pi}{\per\year}$ and $R = \SI{1}{\astronomicalunit}$. Again, we choose to align the waveguide such that the $z$ axis points along the momentary direction of motion. Note however, that if the alignment relative to Earth is kept fixed, then $V^z$ varies over time due to Earth’s spin, so $\delta \beta$ will oscillate with a period of one day.

\begin{table}[h]
	\centering
	\begin{tabular}{@{}llll@{}}
		\toprule
			& Spin (Equator)
			& Spin (Vienna) 
			& Orbit
		\\
		\midrule
		Geometrical Optics
			& \num{1.0526e-6}
			& \num{7.8478e-7}
			& \num{6.7669e-5}
		\\
		Linearisation
			& \num{1.0526e-6}
			& \num{7.8478e-7}
			& \num{6.7669e-5}
		\\
		Exact
			& \num{1.0527e-6}
			& \num{7.8487e-7}
			& \num{6.8408e-5}
		\\
		\bottomrule
	\end{tabular}
	\caption{Numerical results for $\delta \beta/\beta$ due to Earth's rotation.
	Here and in the following tables, “spin” refers to Earth’s rotation about its own axis and “orbit” refers to the rotation about the Sun.}
	\label{table:dispersion rel}
\end{table}

Table \ref{table:dispersion rel} summarises the numerical results for the relative shift $\delta \beta/\beta$. Here, the first estimate was obtained from \eqref{eq:geom:main result}, the “linearisation” result was obtained from \eqref{eq:dispersion:linearised result} and the “exact” number was found by solving \eqref{eq:dispersion:exact} numerically.
The linearised result agrees numerically with the prediction from geometrical optics, which approximates the “exact” result reasonably well.

While these numbers quantify the anisotropy in the propagation of light, one may also compare the wavelength of two parallel light rays propagating at different heights, e.g.\ in a Mach-Zehnder type interferometer whose two arms are displaced by a height difference $\Delta h$.
In view of the high quality of the geometrical optics approximation, we discuss this setup using the formula \eqref{eq:geom:main result}.
If $\vartheta$ denotes the angle between the direction of wave propagation and the local velocity vector field $\vec v$, the geometrical optics formula
\begin{equation}
	\delta \beta = \omega v_\parallel = \omega v \cos \vartheta\,,
\end{equation}
implies a difference in wavelengths in the two arms of
\begin{equation}
	\Delta \beta = \omega \Omega \Delta h \cos \vartheta\,.
\end{equation}
If $\ell$ denotes the length of the interferometer arms, the phase difference at the recombination point evaluates to
\begin{equation}
	\Delta \Phi_\text{rot.} = \omega A \cos \vartheta\,,
\end{equation}
where $A = \ell \Delta h$ is the enclosed area.

We compare this effect with the shift of $\beta$ caused by Earth’s gravitational field.
In \cite{Beig_2018} it was found that a vertical displacement of the dielectric by a heigh $\delta h$ results in the following first order change in the wavelength $\beta$:
\begin{equation}
	\frac{\delta \beta}{\beta}\bigg|_{\mathrm{G}}
	\approx - \delta h \times \SI{2.19e-16}{\per\meter}
	\,.
\end{equation}
Since $-2 \mathrm g/c^2 \approx \SI{2.18e-16}{\per \meter}$, this is in good agreement with the following formula, which we derive in Appendix~\ref{appendix:gravitational field}:
\begin{equation}
	\frac{\delta \beta}{\beta}\bigg|_{\mathrm{G}}
	\approx - 2 \mathrm g \, \delta h/c^2
	\,,
\end{equation}
where $\mathrm g$ is the local gravitational acceleration.
Even for vertical distances of the order of $\SI{100}{\meter}$, the anisotropy effect due to Earth’s spin in Vienna is seven orders of magnitude larger. This suggests that second order effects due to Earth’s spin may still be comparable with the first order effect due to Earth’s gravitational field.
Furthermore, the third order effect due to Earth’s orbital motion might still be larger than the first order gravitational effect. At such a level of accuracy, additional effects stemming from the eccentricity of the orbit might become relevant as well.

In the same Mach-Zehnder setup as described above, the phase shift due to the gravitational potential is found to be given by
\begin{equation}
	\Delta \Phi_\text{grav.} = n \omega g A\,,
\end{equation}
see e.g.~\cite{Hilweg_2017}.
Comparing with the previous formula and restoring the correct powers of $c$, one finds obtains the ratio
\begin{equation}
	\frac{\Delta \Phi_\text{rot.}}{\Delta \Phi_\text{grav.}} = \frac{\Omega c}{n g} \cos \vartheta\,.
\end{equation}
In the considered setup, the maximal ratio of the two phase shifts evaluates to $\Omega c/(ng) \approx \SI{1.6e3}{}$, so the rotational effect is three orders of magnitude larger than the gravitational one.

Let us also note that Earth’s spin causes a height difference of $\Delta h = 2 R_E$ in the gravitational field of the Sun twice a day. Since the Sun has a local acceleration of roughly $\mathrm g_S \approx \SI[per-mode=repeated-symbol]{5.93e-3}{\meter\per\second\squared}$, which – by the above formula – corresponds to a relative shift in $\beta$ of $\SI{-1.68e-12}{}$. Even though this effect is much larger than the gravitational effect discussed in \cite{Beig_2018}, it is unlikely to be directly observable since the effect seems to effect light in all arms of an interferometer in the same way.
\subsection{Main Mode Amplitudes}

Having solved for $\beta$ in terms of $\omega$, the coefficients $\bs \alpha$ must be in the kernel of the continuity matrix for the main mode, as stated in \eqref{eq:continuity main mode}.
The kernel was computed numerically and the spanning vector $\bs \alpha$ was normalised to $\alpha_1 = 1$.

\begin{table}[h]
	\centering
	\begin{tabular}{@{}lSSSS@{}}
		\toprule
			& {Unperturbed}
			& {Spin (Equator)}
			& {Spin (Vienna)}
			& {Orbit}
		\\
		\midrule
		$\alpha_1$
			& 1
			& 1
			& 1
			& 1
		\\
		$\alpha_2$
			& 1.6799
			& 1.6800
			& 1.6800
			& 1.6811
		\\
		$i \alpha_3$
			& 0.6769
			& 0.6765
			& 0.6766
			& 0.6512
		\\
		$i \alpha_4$
			& 1.1454
			& 1.1447
			& 1.1449
			& 1.1027
		\\
		\bottomrule
	\end{tabular}
	\caption{%
	Comparison of the main mode coefficients $\bs\alpha$ in the unperturbed and perturbed case.
	The linearity of the equations was used to normalise the coefficients sucht that $\alpha_1 = 1$.
	“Spin” refers to Earth’s intrinsic rotation and “Orbit” to the rotation around the Sun.
	}
	\label{table: main mode coefficients}
\end{table}

Table~\ref{table: main mode coefficients} shows the numerical results for the coefficients $\bs \alpha$ (normalised to $\alpha_1 = 1$).
As before, the alignment was chosen such that the direction of wave propagation is parallel to the local velocity vector $\vec V$.
We note that the perturbation changes only the modulus of the coefficients, but not their phase.

Comparing, for example, the various results for $\alpha_4$, we find a relative change of $3.7\%$ due to Earth’s orbit around the Sun and a change of only $0.06 \%$ due to its intrinsic rotation at the equator.
This ratio is due to the fact that Earth’s orbital velocity is roughly \num{60} times larger than the tangential velocity at the equator.

\section{Conclusion}

We have computed the first order corrections to electromagnetic waves in cylindrical waveguides due to Earth's rotation by solving Maxwell's equations in the linearised Born metric.
The correction to the dispersion relation was found to be in good agreement with the geometrical optics formula
\begin{equation}
	\frac{\delta \beta}{\beta} = \frac{v_\parallel}{n}
	\,,
\end{equation}
where $n$ is the effective refractive index and $v_\parallel$ is 
the projection of the velocity vector $\vec v$ on the direction of wave propagation.
The relative change of the wave vector $\beta$ due to Earth’s rotation about the Sun was found to be roughly \num[mode=text]{7e-5}, while the effect of Earth’s rotation about its own axis causes relative changes of up to \num[mode=text]{1e-6} at the equator (the effect depends on the latitude and on the orientation of the waveguide in space).
These corrections (relative to the unperturbed value) are many orders of magnitude larger than those due to Earth’s gravitational field.
Furthermore, considering a Mach-Zehnder type interferometer whose arms are placed at different heights, it was shown that the phase shift due to rotation is – depending on the orientation – up to three orders of magnitude larger than the one due to Earth’s gravitational field.

It was found that Earth’s rotation couples neighbouring modes of different angular dependence.
The relative amplitude of the sidebands caused by Earth’s intrinsic rotation was found to be of the order of \num[mode=text]{e-5}, whereas Earth’s orbital rotation around the Sun induces relative amplitudes of the order of \num[mode=text]{e-3}.

We expect that higher order contributions to the dispersion relation cannot be computed using geometrical optics alone, since this method only approximates the result obtained here with an error which we estimate to be much larger than the second order correction.


\appendix
\section{Derivation of the Wave Equation}
\label{a:EM wave equation}

In this section, we derive a wave equation for the electromagnetic field, taking all terms of order $\order(\Omega)$ into account.

We start by writing \eqref{eq:rotation:maxwell covariant} as
\begin{equation}
	\label{eq:wave:maxwell}
	\begin{aligned}
		n^2 \dot {\vec b}
		+ \mu \vec \nabla \times \vec d
		- \nabla_{\vec v}\, \vec b
		+ \vec \Omega \times \vec b
		&= 0 \,,
		\\
		n^2 \dot {\vec d}
		- \epsilon \vec \nabla \times \vec b
		- \nabla_{\vec v}\, \vec d
		+ \vec \Omega \times \vec d
		&= 0 \,,
	\end{aligned}
\end{equation}
which is equivalent to the original form since
\begin{equation}
	[\vec v, \vec w]^i
	= v^j \p_j w^i - w^j \p_j v^i
	= v^j \p_j w^i - w^j \t\epsilon{^j^i^k}\Omega^k
	\,.
\end{equation}
Differentiating the first equation in \eqref{eq:wave:maxwell} with respect to time and multiplying by $n^2$, one obtains
\begin{equation}
	n^4 \ddot{\vec b}
	+ \mu \vec \nabla \times (n^2 \dot{\vec d})
	- n^2 \nabla_{\vec v}\,\dot{\vec b}
	+ \vec \Omega \times (n^2 \dot{\vec b})
	= 0 \,.
\end{equation}
Due to the second equation of \eqref{eq:wave:maxwell} and $n^2 = \epsilon \mu$, the second term takes the form
\ptc{I haven't checked the details of this calculation, but the end result is apparently the same as before? checked previously}
\begin{equation}
	\mu \vec \nabla \times (n^2 \dot{\vec d})
	= n^2 \vec \nabla \times (\vec \nabla \times \vec b)
	+ \mu \vec \nabla \times (\nabla_{\vec v}\, \vec d)
	- \mu \vec \nabla \times (\vec \Omega \times \vec d)
	\,.
\end{equation}
Since $\vec \nabla \cdot \vec b = 0$, the double curl reduces to
\begin{equation}
	\vec \nabla \times (\vec \nabla \times \vec b)
	= \vec \nabla (\vec \nabla \cdot \vec b) - \Delta \vec b
	= - \Delta \vec b
	\,,
\end{equation}
where $\Delta \vec b$ denotes the vector Laplacian of $\vec b$.
Moreover, due to $\vec \nabla \cdot \vec d = 0$ and $\vec v = \vec \Omega \times \vec r$, we have
\begin{equation}
	\vec \nabla \times (\vec \nabla_{\vec v}\, \vec d)
	= \nabla_{\vec v}\, (\vec \nabla \times \vec d)
	- \vec \nabla (\vec \Omega \cdot \vec d)
	\,.
\end{equation}
Together with the intermediate result
\begin{equation}
	\vec \nabla \times (\vec \Omega \times \vec d) = - (\vec \Omega \cdot \vec \nabla) \vec d
	\,,
\end{equation}
which is also due to the fact that $\vec \nabla \cdot \vec d = 0$, we obtain
\begin{equation}
	\begin{aligned}
		\mu \vec \nabla \times (n^2 \dot{\vec d})
		&= - n^2 \Delta \vec b
		+ \nabla_{\vec v}\, (\mu  \vec \nabla \times \vec d)
		- \mu \vec \nabla (\vec \Omega \cdot \vec d)
		+ \mu (\vec \Omega \cdot \vec \nabla) \vec d
		\\
		&= - n^2 \Delta \vec b
		- n^2 \nabla_{\vec v}\, \dot {\vec b}
		- \mu [
			\vec \nabla (\vec \Omega \cdot \vec d)
			- (\vec \Omega \cdot \vec \nabla) \vec d
			\,
		]
		+ \order(\Omega^2)
		\,.
	\end{aligned}
\end{equation}
Combining this with
\begin{equation}
	n^2 \dot{\vec b} \times \vec \Omega
	= \mu \vec \Omega \times (\vec \nabla \times \vec d) + \order(\Omega^2)
	= \mu [
			\vec \nabla (\vec \Omega \cdot \vec d)
			- (\vec \Omega \cdot \vec \nabla) \vec d
			\,
		]
		+ \order(\Omega^2)
	\,,
\end{equation}
we arrive at the wave equation
\ptc{this one is apparently more convenient for some calculations}
\begin{equation}
	n^2 (
		n^2 \ddot{\vec b}
		- \Delta \vec b
	)
	= 2 \nabla_{\vec v}\, (n^2 \dot{\vec b})
	- 2 \mu [
		(\vec \Omega \cdot \vec \nabla) \vec d - \vec \nabla (\vec \Omega \cdot \vec d)
	]
	+ \order(\Omega^2)
	\,.
\end{equation}
%

\section{Remarks on the Influence of Earth’s Gravitational Field}
\label{appendix:gravitational field}

We give a derivation of the first order change in the wave vector due to Earth’s gravitational field using geometrical optics.

We start from the post-Newtonian metric in the form
\begin{equation}
	\t g{_\mu_\nu}
	= \t \eta{_\mu_\nu} - 2 \phi\, \t \delta{_\mu_\nu}
	\,,
\end{equation}
where $\phi$ is Newton’s potential.
To first order in $\phi$, the components of the inverse metric tensor read
\begin{equation}
	\t g{^\mu^\nu}
	= \t \eta{^\mu^\nu} + 2 \phi\, \t \delta{^\mu^\nu}
	\,.
\end{equation}
A medium at rest in the considered coordinates has four velocity $(u^\mu) = (u^0, \vec 0)$, where $u^0$ is determined from $g(u,u) = -1$, which yields
\begin{equation}
	(u^0)^2 = - \frac{1}{g_{00}} \approx 1 - 2 \phi\,.
\end{equation}
Inserting this into \eqref{eq:optical metric inverse components}, one obtains the components of the inverse optical metric 
\begin{equation}
	\begin{aligned}
		\t \go{^0^0} &= -1 + 2 \phi + (1-n^2)(1- 2 \phi) = -n^2 (1 - 2 \phi)\,,\\
		\t \go{^i^j} &= (1 + 2 \phi) \t \delta{^i^j}\,,\\
		\t \go{^0^i} &= \t \go{^i^0} = 0 \,.
	\end{aligned}
\end{equation}
The condition of $k = \omega \dd t - \beta \dd z$ being null with respect to $\go$ then yields the quadratic equation
\begin{equation}
	(1+2\phi) \beta^2 - n^2 (1 - 2 \phi)\omega^2 = 0\,,
\end{equation}
whose solution for positive $\beta$ ($\omega > 0$) is
\begin{equation}
	\beta = n \omega \sqrt{\frac{1 - 2 \phi}{1+2 \phi}} \approx n \omega (1 - 2 \phi) \,.
\end{equation}
The linearised change in $\beta$ due to the gravitational field is thus
\begin{equation}
	\frac{\delta \beta}{\beta} = - 2 n \frac{\omega}{\beta} \delta \phi\,,
\end{equation}
or, using $\beta = n \omega + \order(\phi)$ and $\delta \phi = \mathrm g \delta h$, where $\mathrm g$ is the local gravitational acceleration and $\delta h$ is a difference in height
\begin{equation}
	\frac{\delta \beta}{\beta}
	= - 2 \mathrm g \delta h \,.
\end{equation}
This is the same result as equation~(1.1) in \cite{Beig_2018}.

\clearpage
\addcontentsline{toc}{section}{References}
\bibliographystyle{plain}
\bibliography{bibliography}

\end{document}